# Structural, magnetic, and optical properties of zinc- and copper-substituted nickel ferrite nanocrystals

F. Shahbaz Tehrani [a]; V. Daadmehr [a,*]; A. T. Rezakhani [b]; R. Hosseini Akbarnejad [a]; S. Gholipour [a]

[a] *Magnet& Superconducting Lab, Department of Physics, Alzhara University, Tehran 19938, Iran*

[b] *Department of Physics, Sharif University of Technology, Tehran, Iran*

## Abstract

Ferrite nanocrystals are interesting material due to their rich physical properties. Here we add nonmagnetic dopants Zn and Cu to nickel ferrite nanocrystals, $Ni_{1-x}M_xFe_2O_4$ ($0\leq x\leq 1$, $M\in\{Cu, Zn\}$), and characterize how relevant properties of the samples are modified accordingly. Basically, these dopings cause a rearrangement of $Fe^{+3}$ ions into the two preexisting octahedral and tetrahedral sites. In fact, this, we show, induces pertinent magnetic properties of the doped samples. In the case of the Cu-doping, the Jahn-Teller effect also emerges, which we identify through the FTIR Spectroscopy of the samples. Moreover, we show an increase in the lattice parameters of the doped samples, as well a superparamagnetic behavior for the doped samples is shown, while the Jahn-Teller effect precludes a similar behavior in the $CuFe_2O_4$ nanocrystals. The influences of Zn and Cu substitutions are investigated on the optical properties of nickel ferrite nanocrystals by photoluminescence measurement at room temperature.

## 1. Introduction

Spinel ferrites, with common formula of $MFe_2O_4$ (M: a divalent metal ion), have wide technological applications, e.g., in multilayer chip inductor (MLCI), ferrofluids, high-speed digital tape or recording disks, rod antenna, and humidity sensor [1-9]. Ferrite nanocrystals are also of interest in various applications, such as inter-body drug delivery [10-12], bioseparation, and magnetic refrigeration systems [13], in particular due to their specific properties, such as superparamagnetism. In addition, among ferrospinels zinc ferrites are used in gas sensing [14, 15], catalytic application [16], photocatalyst [17, 18], and absorbent materials [19].

The unit cell of spinel ferrites is composed of 32 oxygen atoms in cubic closed- packed arrangement distributed in tetrahedral ('A') and octahedral sites ('B'). Chemical and structural properties of spinel ferrite nanocrystals are affected by their compositions and synthesis methods, and corresponding electric and magnetic properties depends on cation substitutions.

Doping ferrite nanocrystals with various metals, such as chromium, copper, manganese, and zinc are usually used to improve some of their electric or magnetic properties [20-22]. For example, Zn/Ni ferrites have applications as soft magnetic materials with high frequency (due to high electrical resistivity and low eddy-current loss [8]). Along that line, (Cu, Zn)/Ni ferrites offer a further improvement as softer magnetic materials [23].

In this work, we study effects of Zn/Ni and Cu/Ni substitutions on nickel ferrite nanocrystals synthesized through the sol-gel method. Specifically, we characterize structural and magnetic properties of the doped samples at room temperature. We also exhibit the emergence of the Jahn-Teller effect in the case of the $CuFe_2O_4$ nanocrystals, and argue that the modification of the structural properties is in principle due to the occupation of the A sites vs. B sites. In this exhaustive characterization, various techniques are employed such as X-ray diffraction (refined with the Rietveld method with the MAUD software), Field-Emission Scanning Electron Microscope (FE-SEM), Fourier Transform Infra-Red (FTIR) spectroscopy, Vibrating Sample Magnetometer (VSM), and Photoluminescence (PL), and UV-Vis spectrometer at room temperature. We thus compare the effect of dopings not only regarding structural and magnetic properties, but also in the sense of optical properties. In addition, here the Jahn-Teller effect has is studied through the investigation of the structural and magnetic properties.

## 2. Experimental details

### *2.1. Preparation method*

The sol-gel method is widely used in synthesis of ferrite nanocrystals because of its high reaction rate, low preparation temperature, and production of small particles. Hence, in our experiment, magnetic nanocrystalline $Ni_{1-x}Zn_xFe_2O_4$ with x=0, 0.3, 0.5, 0.7, 1 and $Ni_{1-x}Cu_xFe_2O_4$ with x=0, 0.5, 1 were synthesized by this method. Citric acid $C_6H_8O_7$, ferric nitrate $Fe(NO_3)_3.9H_2O$ (98%), nickel nitrate $Ni(NO_3)_2.6H_2O$ (99%), zinc nitrate $Zn(NO_3)_2.6H_2O$ (99%) and copper nitrate $Cu(NO_3)_2.3H_2O$ (99%) were produced by Merck® company and were used as raw materials. The stoichiometric amount of nitrates and acid citric were dissolved separately in deionized water to make 0.5M solutions. The mole ratio of metal nitrates to citric acid was taken as 1:1. To obtain smaller nanocrystals, we added ethylenediamine to the solution until its pH became 1. Next, the sol was heated continuously at 70°C under stirring to form a brown dried gel. This gel was fired at 135°C in oven for 24h, and was then ground into powder. To obtain various $Cu^{+2}$ and $Zn^{+2}$ substitutions, we calcined the powder at 300-600°C for 4h.

*2.2. Measurements and characterizations*

The X-ray diffraction patterns (XRD) of the synthesized nanocrystals were obtained using a Philips® PW1800 X-ray diffractometer with Cu Kα radiation (λ= 1.54056Å) operated at 40kV and 30mA. The refinement method of Rietveld was applied with the "Material Analysis Using Diffraction" (MAUD) program (v.2.056). Fourier Transform Infra-Red (FTIR) spectra of samples were detected by a BRUKER®TENSOR27 FTIR spectrometer with transmission from 4000 to 400cm$^{-1}$ using KBr pellets. The average grain size and morphology of the samples were observed by a Hitachi® S4160 Field Emission Scanning Electron Microscopy (FE-SEM). Photoluminescence (PL) measurement was performed by a VARIAN® CARY ECLIPSE spectrometer (using a diode laser with the wavelength of 430nm). The optical absorption spectra were recorded by a Perkin-Elmer® Lambada35 UV-Vis spectrometer. The magnetic properties of the samples were measured by Meghnatis Daghigh Kavir Co.® Vibrating Sample Magnetometer (VSM) at room temperature. Note that during such measurements the maximum applied magnetic field was 9kOe.

## 3. Results and discussion

*3.1. Structural studies*

The X-ray diffraction patterns of the synthesized ferrite nanocrystals have been shown in Figs. 1 and 2. The existence of the (220), (311), (400), (422), (511) and (440) major lattice planes in the XRD patterns confirms the formation of spinel cubic structure with the Fd3m space group, which is consistent with the powder diffraction file of JCPDS. Also the presence of the (111), (222), (331), (533), (622), (444), (642) and (731) minor lattice planes in the XRD patterns agrees well with the powder diffraction of spinel cubic JCPDS file. All samples are considered to be single-phase spinel structure. In addition, as Fig. 2 shows, the XRD patterns of the $Ni_{0.5}Cu_{0.5}Fe_2O_4$ and $CuFe_2O_4$ nanocrystals confirm that almost all Cu atoms have been placed in the cubic spinel structure. However, note that in some previous studies, there are reports of excess phases, e.g., the CuO phase [24] or tetrahedral $CuFe_2O_4$ phase [25]. Specifically, it has been reported in Ref. [24] that both cubic spinel structure and CuO phase were observed in the $Ni_{0.5}Cu_{0.5}Fe_2O_4$ nanocrystals (with pH 3-4) when the calcination process lasted for 3h at 600°C, while only the single cubic spinel structure was formed when the calcination temperature was up to 1000°C. In the case of our experiment, the $Ni_{0.5}Cu_{0.5}Fe_2O_4$ nanocrystals (with pH=1) were calcined at 375°C for 4h, and crystallized in the single cubic spinel structure with almost no excess phases.

The average crystallite sizes D are calculated from the characteristics of the (311) XRD-peaks through the Scherrer formula in the range of 1.4-19.6nm (Tables 1 and 2). Note that, similar to the case of Zn/Ni ferrites prepared by the combustion-reaction method [26], here an increase in the Zn and Cu contents does in fact increase the particle size. The existence of broad peaks in the XRD pattern of our synthesized nickel ferrite nanocrystals can be attributed to the small crystallite size of them (see the inset of Fig. 1).

The X-ray diffraction data of the synthesized ferrite nanocrystals has been refined by using the MAUD software and Rietveld's method for the structural analysis, cation distribution and lattice parameter calculations. The obtained lattice parameters for the nanocrystals are listed in Tables 1 and 2. It is seen that the lattice parameters increase for larger Zn and Cu contents. This increase can be attributed to larger ionic radiuses of $Cu^{2+}$ (0.72Å) and $Zn^{2+}$ (0.82Å) relative to $Ni^{2+}$ (0.69Å); this is consistent with Refs. [24,27]. In addition, because the ionic radius of $Zn^{2+}$ is larger than the ionic radius of $Cu^{2+}$, the $Zn^{2+}$ substitution leads to larger expansion of the lattice; thus the lattice parameter increases more in comparison to the $Cu^{2+}$ substitution in the synthesized nanocrystals. Also, the shift of the (311) XRD peak to small diffraction angle with the increase in the Zn and Cu contents may be attributed to this point that the sample with higher Zn and Cu contents has a larger lattice parameter. These shifts are larger for the $Ni_{1-x}Zn_xFe_2O_4$ nanocrystals because of larger ionic radius of $Zn^{2+}$ in comparison to $Cu^{2+}$.

To ensure that spinel structure for the synthesized nanocrystals has been formed, as well to investigate how the ferromagnetic ions ($Fe^{3+}$ ,$Ni^{2+}$) and nonmagnetic transition metal ions ($Cu^{2+}$ ,$Zn^{2+}$) occupy tetrahedral and octahedral sites, we refined the XRD data by employing the MAUD software. Table 3 shows the before- and after-refinement values, indicating relatively high accuracy in synthesizing the nanocrystals without formation of extra phases.

Based on the XRD data refinement, the formation of single-phase spinel cubic structure with Fd3m space group has been confirmed in all samples. Additionally, these results indicate that the synthesized nickel ferrite ($NiFe_2O_4$) and copper ferrite ($CuFe_2O_4$) nanocrystals have inverse spinel structure in which half of the $Fe^{3+}$ ions spatially fill the tetrahedral sites and the rest occupy the octahedral sites with the $Ni^{2+}$ ions in nickel ferrite and the $Cu^{2+}$ ions in copper ferrite nanocrystals. Generally, an inverse spinel ferrite can be represented by the formula $[Fe^{3+}]_{tet}[A^{2+}, Fe^{3+}]_{oct}O_4^{2-}$ (A= Ni, Cu,..), where the "tet" and "oct" indices represent the tetrahedral and octahedral sites, respectively. Likewise, these results specify that the synthesized zinc ferrite ($ZnFe_2O_4$) nanocrystals have a normal spinel structure in which all $Zn^{2+}$ ions fill tetrahedral sites, hence the $Fe^{3+}$ ions are forced to occupy all of the octahedral sites. As a result, this compound can be displayed by the formula $[Zn^{2+}]_{tet}[Fe_2^{3+}]_{oct}O_4^{2-}$.

Also these results indicate that the synthesized Zn/Ni ferrite (0<x<1) and Cu/Ni ferrite ($Ni_{0.5}Cu_{0.5}F_2O_4$) nanocrystals have a mixed spinel structure where the $M^{2+}$ (M= Zn, Cu) cations are substituted for the $Fe^{3+}$ ions in the tetrahedral sites, and these $Fe^{3+}$ ions move to the octahedral sites, as shown by the gray color in Table 3. In other words, in these nanocrystals the $M^{2+}$ (M= Zn, Cu) ions occupy the tetrahedral sites with the $Fe^{3+}$ ions and the while the rest of the $Fe^{3+}$ ions fill the octahedral sites with the $Ni^{2+}$ ions. Thus, these samples can be represented by the formula $[M_x^{2+} ,Fe_{1-x}^{3+}]_{tet} [Ni_{1-x}^{2+} ,Fe_{1+x}^{3+}]_{oct} O_4^{2-}$. All of these results for the structural analysis and cation distribution calculations for synthesized Zn/Ni ferrites are in agreement with the structure and cation distribution for bulk spinel ferrites of Refs. [28,29]. Table 3 represents the occupancy of the $Zn^{2+}$ and $Cu^{2+}$ (for the $Ni_{0.5}Cu_{0.5}F_2O_4$ nanocrystals) contents in the tetrahedral sites, formation of the inverse spinel structure for nickel and copper ferrites, formation of normal spinel structure for zinc ferrites, and formation of mixed spinel structure for the other nanocrystals. Also the cation distribution of the prepared nanocrystals are summarized in Tables 1 and 2, and Fig. 3 represents the XRD patterns for the $ZnFe_2O_4$ nanocrystals.

The pattern fitness can be checked for XRD data. There are several parameters for the calculation of pattern fitness. The goodness of fit (S) is described by $S= R_{wp} /R_{exp}$, where $R_{wp}$ is the weighted residual error and $R_{exp}$ is the expected error. Refinement has been continued until a convergence was reached for a value of S close to 1. which confirms the goodness of refinement. These parameters are listed in Table 4.

For Cu doped samples, we have also tested a mixed scenario that puts the copper into both sublattices (for example the cation distribution for $CuFe_2O_4$ nanocrystals can be displayed by the formula $[Cu_y^{2+}, Fe_{1-y}^{3+}]_{tet}$ $[Cu_{1-y}^{2+}, Fe_{1+y}^{3+}]_{oct}$ , where $0 \leq y \leq 1$ with step 0.1) and thus refined the distribution of the copper in the tetrahedral and octahedral sites and also the distribution and the positions of the other ions by using the MAUD software. With refinement of the XRD data for our Cu doped samples, the MAUD software obtained that for $0<y\leq 1$, it is unable to solve the refinement with the supplied initial parameters, because the correlation matrix from the Choleski decomposition has negative diagonal. Thus we found that for our synthesized $Ni_{0.5}Cu_{0.5}Fe_2O_4$ nanocrystals, all copper ions go into the tetrahedral sites so that the cation distribution for this sample can be displayed by the formula $[Cu_{0.5}^{2+}, Fe_{0.5}^{3+}]_{tet}$ $[Ni_{0.5}^{2+}, Fe_{1.5}^{3+}]_{oct}$. Also for our synthesized $CuFe_2O_4$ nanocrystals, all copper ions go into the octahedral sites thus the cation distribution can be displayed by the formula $[Fe^{3+}]_{tet}$ $[Cu^{2+}$ ,

$Fe^{3+}]_{oct}$. Note that the FTIR results are consistent with the MAUD analysis for the cation distribution. In addition, this cation distribution for Cu-doped samples explains well thier magnetic properties.

*3.2. FTIR study*

The FTIR spectra of the synthesized nanocrystals, shown in Figs. 4 (a) and (b), confirm the formation of the spinel structures suggested by the MAUD analysis. The bands around 3415 and 1621cm$^{-1}$ are due to the O-H stretching vibration of the free or absorbed water. This indicates existence of hydroxyl groups in our synthesized ferrites, which is apparently seen in earlier experiments too [30, 31]. The band at 1380cm$^{-1}$ is attributed to the C=O stretching vibration of the carboxylate group ($CO^{2-}$), and the band around 1099cm$^{-1}$ is related to the NO-stretching vibration due to nitrate group [32, 33]. These bands disappear when Cu and Zn contents (and thus calcination temperature) increased.

In the range of 800-400cm$^{-1}$, two main absorption bands were observed in the range of 604-542cm$^{-1}$ and 463-416cm$^{-1}$. These bands are attributed to the vibration of the tetrahedral and octahedral metal-oxygen (M-O) bonds in the lattices of the synthesized nanocrystals, respectively. These FTIR frequency bands for various Zn and Cu contents have been listed in Tables 5 and 6. It is understood from these tables that the value of $\upsilon_1$ and $\upsilon_2$ are found to decrease and increase, respectively, with the increase in the Zn and Cu contents except for the $CuFe_2O_4$ sample.

These results can be explained as fllows. The addition of $Zn^{2+}$ and $Cu^{2+}$ (from x=0 to x=0.5 for Ni/Cu ferrites) ions in the tetrahedral sites with larger radius and greater atomic weight, makes the $Fe^{3+}$ ions migrate to the octahedral sites, and consequently decreases the tetrahedral vibration frequency. Similarly, migration of the $Fe^{3+}$ ions to the octahedral site increases the octahedral vibration frequency. For the $CuFe_2O_4$ nanocrystals, the increase of the tetrahedral vibration frequency and the decrease of the octahedral vibration frequency may be due to the presence of half of the $Fe^{3+}$ ions in the A sites and presence of the $Cu^{2+}$ ions with larger radius and greater atomic weight in the B sites. There is a weak splitting around octahedral vibration frequency ($\upsilon'_2$) of the $CuFe_2O_4$ nanocrystals that can be attributed to the existence of $Cu^{2+}$ ions, a Jahn-Teller ion [34, 35], in the octahedral sites, and causes some lattice distortion. This distortion normally takes the form of stretching the bonds to the ligand lying along the *z* axis, but sometimes happens as a shortening of these bonds as well. This affects the octahedral Cu-O vibration frequency in the $CuFe_2O_4$ nanocrystals. Note that all of these results are also consistent with the MAUD analysis for the cation distribution.

### *3.3. Morphological study*

The FE-SEM images of the $CuFe_2O_4$, $Ni_{0.5}Zn_{0.5}Fe_2O_4$ and $ZnFe_2O_4$ nanocrystals have been shown in Figs. 5(a)-(c). The average grain size of the $CuFe_2O_4$, $Ni_{0.5}Zn_{0.5}Fe_2O_4$ and $ZnFe_2O_4$ nanocrystals are 15, 20, and 30nm, respectively. Note that the average grain size of the samples obtained from SEM images is larger than nanocrystals size as calculated using the XRD measurements, which simply indicates that each grain is formed by aggregation of a number of nanocrystals. The samples are spherical and uniform, and cohesion of grains is due to the magnetic attraction.

### *3.4. Magnetic study*

The magnetic properties of the synthesized nanocrystals are analyzed using a Magnetometer (VSM) at room temperature. Figures 6(a) and (b) show the M-H curves of the prepared $Ni_{1-x}M_xFe_2O_4$ nanocrystals. The saturation magnetization ($M_s$) and coercivity ($H_c$) values have been directly extracted from these curves and have been listed for various Zn and Cu contents in Tables 1 and 2. These Tables indicate that $M_s$ increases while increasing the Zn/Ni and Cu/Ni contents up to x=0.5 (after which $M_s$ decreases); i.e., the maximum value of $M_s$ is found in the $Ni_{0.5}M_{0.5}Fe_2O_4$ (M=Zn, Cu) nanocrystals. The latter increase is higher in the Zn/Ni substitution against the Cu/Ni substitution.

The saturation magnetization of the synthesized nickel ferrite and zinc ferrite nanocrystals have been obtained at room temperature from the hysteresis loops as 6.4 and 1.7emu/g, respectively (Tables 1 and 2). These values are less than the corresponding values for bulk nickel and zinc ferrites (50 and 5emu/g, respectively) [36, 37]. This observation can be attributed to the presence of small nanoparticles having core-shell morphology, containing a spin-glass-like surface layer and ferrimagnetically lined-up core spins. As a result, the canting and disorder of the surface layer spins may be due to broken superexchange bonds, and unlike local symmetry for those atoms near the surface layer lead to a smaller $M_s$ in these nanocrystals. These effects are especially dominant in the case of ferrites due to the presence of superexchange interactions. In smaller ferrite nanoparticles with larger surface/volume ratio, these effects are strongest. The $M_S$ of the synthesized $NiFe_2O_4$ nanocrystals (with 1.4nm crystallite size) is very smaller than the value for the bulk.

According to Neel's ferrimagnetic theory, in the spinel structure the cations on different sublattices (tetrahedral and octahedral sites) have oppositely aligned magnetic moments [31, 38]. Hence, the magnetic moment per formula unit ($n_B$) in the $\mu_B$ units is:

$$n_B = M_{oct} - M_{tet} \qquad (1)$$

where, $M_{oct}$ and $M_{tet}$ are the magnetic moments of the octahedral and tetrahedral sites, respectively. The magnetic behavior of our synthesized nanocrystals measured by VSM can be attributed to the competition of ferromagnetic ions such as $Fe^{3+}$, $Ni^{2+}$ and the $Cu^{2+}$ and $Zn^{2+}$ ions as non-magnetic transition metal ions in the occupancy of the tetrahedral and octahedral sites. This can be explained by the obtained cation distribution from the MAUD analysis of XRD patterns (summarized in Tables 1 and 2), and Neel's theory with the knowledge of ionic magnetic moment of 0, 1, 2, and 5$\mu_B$ for $Zn^{2+}$, $Cu^{2+}$, $Ni^{2+}$ and $Fe^{3+}$ ions, respectively.

It is evident from Tables 1 and 2 that the saturation magnetization of the synthesized nanocrystals has been enhanced with the increase in Zn and Cu contents up to x=0.5. This is consistent with the obtained cation distribution the MAUD analysis of the XRD patterns. The occupancy of $Zn^{2+}$ and $Cu^{2+}$ substituted ions in the tetrahedral sites causes transfer of $Fe^{3+}$ ions from there to the octahedral sites. This increases the concentration of $Fe^{3+}$ ions in octahedral sites and so increases the magnetization in the octahedral sublattice. This in turn enhances the magnetization of the nanocrystals. These results agree well with the bulk sample in that the maximum $M_s$ is found in the $Ni_{0.5}Zn_{0.5}Fe_2O_4$ for Zn/Ni bulk ferrites. Tables 1 and 2 indicate that the increase in $M_s$ for Zn/Ni ferrite is more than Cu/Ni ferrite nanocrystals. This can be explained by the total spin moment of the tetrahedral site and octahedral site interactions. Due to placement of $Zn^{2+}$ and $Cu^{2+}$ ions in the tetrahedral sites, the magnetization of the B sublattice is expected to be identical for $Ni_{0.5}Zn_{0.5}Fe_2O_4$ and $Ni_{0.5}Cu_{0.5}Fe_2O_4$ nanocrystals, while for the A sublattice it increase with substitution of higher magnetic moment of $Cu^{2+}$ ($1\mu_B$) in comparison to $Zn^{2+}$ ($0\mu_B$). Consequently, because of the antiferromagnetic coupling, the result is a decrease in the net overall magnetic moment (Eq.1) of the $Ni_{0.5}Cu_{0.5}Fe_2O_4$ relative to the $Ni_{0.5}Zn_{0.5}Fe_2O_4$ nanocrystals. In addition, higher calcination temperature for the $Ni_{0.5}Zn_{0.5}Fe_2O_4$ (450°C) in contrast to the $Ni_{0.5}Cu_{0.5}Fe_2O_4$ (375°C) nanocrystals implies larger grain size and higher crystalline quality in the ferrites, thereby leads to higher $M_s$ value.

It is observed from Tables 1 and 2 that with the increase in the Zn and Cu contents from x=0.5 to x=1, $M_S$ shows a decrease. This reduction is remarkable for the Zn/Ni substitution. The decrease of $M_S$ for the $Ni_{0.3}Zn_{0.7}Fe_2O_4$ nanocrystals is referred to as spin canting in the surface layer of nanoparticles. When the Zn content increases to x=0.7, the occupation of $Zn^{2+}$ ions in the tetrahedral sites decreases the number of spins occupation in the A sites, thus the A-B superexchange interactions are weakened and the spines of the octahedral sites cannot hold collinearity to the little tetrahedral spins. Therefore, the $M_s$ value decreases.

Based on the MAUD analysis and the VSM measurements in Fig. 6 (a), the small $M_s$ of $ZnFe_2O_4$ can be attributed to the absence of the $Ni^{2+}$ ions as ferromagnetic ions and the occupancy of $Zn^{2+}$ ions with zero magnetic moment in all tetrahedral sites, i.e., replacing the complete $Fe^{3+}$ ions in the octahedral sites. Consequently, the interactions between $Fe^{3+}$ ions in the octahedral sites are too weak to be important (the B-B interaction is weak in ferrites), hence the magnetization of $ZnFe_2O_4$ nanocrystals is small. Also, with the increase in the Cu content from x=0.5 to x=1, a small reduction is observed in the saturation magnetization. To explain this behavior, we can argue that based on the MAUD analysis, the $Cu^{2+}$ ions move from the tetrahedral to octahedral sites, which have smaller ionic magnetic moment in contrast with the $Ni^{2+}$ ions as mentioned above, and the half of $Fe^{3+}$ ions fill the A sites, so reduce the superexchange interactions between octahedral and tetrahedral sites, which cause the decrease in the $M_s$ value. But since the Jahn-Teller effect occurs, this decrease is small. The existence of $Cu^{2+}$ ion as a Jahn-Teller ion in the octahedral sites of the $CuFe_2O_4$ nanocrystals causes a lattice distortion, which has the effect of removing the electronic and orbital degeneracies of $Cu^{2+}$ cations. It is anticipated that this effect in turn creates large strains in the copper ferrite lattice, and as a result modified magnetic properties. Hence, this causes a reduction in saturation magnetization from x=0.5 to x=1 to be smaller. Note that the Jahn-Teller theorem does not predict the direction of the distortion; it only marks the existence of unstable lattice geometry.

Since the coercivity ($H_c$) of a magnetic material is a measure of its magneto-crystalline anisotropy, the small nanoparticles, which have close to zero coercivity and no remanence become single domain with little anisotropy energy. This is a characteristic of superparamagnetic nanocrystals. The coercivity ($H_c$) of the synthesized nanocrystals has been obtained from Figs.6 (a) and (b), and is listed in Tables 1 and 2. Note that the $H_c$ values of all prepared nanocrystals are close to zero, except $CuFe_2O_4$ nanocrystals (Tables 1 and 2). Thus these nanocrystals show superparamagnetic behavior at room temperature. Such properties make these materials favorable for wide engineering applications such as drug delivery [39], bioseparation [40] and magnetic refrigeration systems [41].

The $H_c$ of the samples increases with the Zn/Ni and Cu/Ni substitutions. This increase is small for Zn/Ni ferrite nanocrystals, while it is significant for Cu/Ni ferrite nanocrystals. This increase can be attributed to higher magneto-crystalline anisotropy of $Cu^{2+}$ in comparison to $Ni^{2+}$ and $Zn^{2+}$ ions (hence the Jahn-Teller effect). Although the $CuFe_2O_4$ nanocrystals have average crystallite size smaller than 10nm, they have large coercivity, as a result of complete substitution of the $Cu^{2+}$ to the $Ni^{2+}$ ions in the octahedral sites and the occurrence of the Jahn-Teller effect. The existence of $Cu^{2+}$ as a Jahn-Teller ion in the octahedral sites of the $CuFe_2O_4$ nanocrystals causes the lattice distortion which is anticipated to create large strains in the copper ferrite lattice. This increases the anisotropy and coercivity of this sample, which agrees well with the FTIR analysis of the $CuFe_2O_4$ nanocrystals (see Table 6). Likewise, this effect shows its small influence on the coercivity in $Ni_{0.5}Cu_{0.5}Fe_2O_4$ as a result of that the $Cu^{2+}$ ions fill the tetrahedral sites. Thus the Jahn-Teller effect is much less noticeable in tetrahedral complexes.

*3.5. Optical properties*

Photoluminescence spectroscopy is measured at room temperature to detect the optical properties of all synthesized nanocrystals for various substitutions of the $Zn^{2+}$ and $Cu^{2+}$ ions (Figs. 7(a) and 7(b)). Five clear emission bands have been observed in the entire PL spectra for these nanocrystals. The emission bands, around 541.94 and 518.93 and 460.89nm are constant with the increase in the Zn and Cu content, and are attributed to the $3d^5 \rightarrow 3d^3\ 4s^2$ transitions of $Fe^{3+}$ ions. We remark that in Ref. [42] similar transitions at 530 and 470nm have been reported for $ZnFe_2O_4$ thin films.

In addition, the intensity of the emission bands at 541.94 and 518.93nm vary with the increase in the Zn/Ni substitution. The dependence of these transition intensities of $Fe^{3+}$ ions on the Zn content (x) is depicted in Fig. 8. It is evident that these intensities increase with the increase in the Zn content except for x=0.5. This may be explained as follows: i) Based on the MAUD analysis, with the increase in the $Zn^{2+}$ concentration and leaving of the $Ni^{2+}$ cations in the structure, $Zn^{2+}$ ions occupy the tetrahedral sites and transfer the $Fe^{3+}$ to the octahedral sites. ii) These changes cause a decrease in the structural isotropy of the synthesized nanocrystals except for the $Ni_{0.5}Zn_{0.5}Fe_2O_4$ nanocrystals. iii) The nanocrystals have biggest saturation magnetization among the synthesized nanocrystals and have a structural isotropy in the tetrahedral sites. iv) The occupation of half of the tetrahedral sites by the $Zn^{2+}$ ions and the rest by the $Fe^{3+}$ ions causes the decrease in transitions of the $Fe^{3+}$ ions in the tetrahedral sites.

The broad emission bands around 660 and 484nm are attributed to the indirect and direct band gaps for our synthesized nanocrystals, respectively. The values of the indirect and direct band gaps are listed for various substitutions of the Zn and Cu content in Tables 5 and 6. In particular, there are two minor peaks detected by the spectrometer around 660nm and 484nm, presenting, respectively, the indirect and direct band gaps ($E=hc/\lambda$). Admittedly the accuracy of this estimation is not high. The minor peaks around 660nm for $Ni_{1-x}Cu_xFe_2O_4$ nanocrystals are displayed next to Fig. 7 (b). However, to confirm our results, we also performed a UV-Vis experiment for all samples. Figure 9 shows the optical absorption spectra for our $Ni_{1-x}Zn_xFe_2O_4$ nanocrystals in the UV region. The optical band gap $E_g$ can be determined by the equation $\alpha E_p=A(E_p-E_g)^q$ [43], where $E_p$ is the photon energy, A is a constant that depends on the transition probability and q depends on the nature of the transition and is theoretically equal to 2 and 1/2 for allowed indirect and direct electronic transition, respectively. The connection of the linear fits of $(\alpha E_p)^{1/q}$ versus $E_p$ plots for q=1/2 and 2 (shown in Fig. 10) on the $E_p$-axis determines the direct and indirect band gaps, respectively. The obtained indirect and direct band gaps from the UV-Vis spectra are consistent with these results from the PL measurements.

Note that the value of the indirect band gap increases with the increase in the Zn content, while it is almost constant with the Cu content. The values of the direct band gap are constant for various substitutions of Zn and Cu content. The value of the indirect band gap is affected by various factors such as crystallite size, structural parameter, and presence of impurities. The increase in the indirect band gap for our synthesized $Ni_{1-x}Zn_xFe_2O_4$ nanocrystals may be attributed to the more increase in the structural parameter (lattice constant) with the increase in the Zn concentration in comparison to the Cu content (Tables 1 and 2).

The obtained indirect band gap values for our synthesized Zn/Ni nanocrystals (1.85–1.90eV) are higher than the determinant values for the Zn/Ni ferrite nanocomposites (1.50–1.66eV) in Ref. [44], which is due to presence of impurities. Note that the obtained values of the indirect and direct band gaps for our prepared $ZnFe_2O_4$ nanocrystals are higher than estimated values of the direct and indirect band gaps for zinc ferrite nanowires [45], which are 2.23 and 1.73eV, respectively and are similar to the $ZnFe_2O_4$ films [46].

## 4. Conclusion

We have investigated experimentally how properties of spinel nickel ferrites are affected because of doping by Cu and Zn. We have prepared these samples through the sol-gel method, and performed various structural, magnetic, and optical measurements on them. The crystallite size of synthesized nanocrystals was calculated using Scherrer's formula in the range of 1.4-19.6nm. The X-ray diffraction measurements, refined with the Rietveld method, have indicated formation of a single-phase cubic spinel with the space group Fd3m, and cation distributions. In particular, the analysis of X-ray diffraction has showed that the structural and magnetic properties are induced by competition of ferromagnetic ions $Ni^{+2}$ and $Fe^{+3}$ and nonmagnetic ions $Zn^{+2}$ and $Cu^{+2}$ ions in occupation of the tetrahedral and octahedral sites. Moreover it has been shown that, as in the bulk case, $Zn^{+2}$ cations have preferred to substitute for $Fe^{+3}$ ions in the tetrahedral sites. Magnetic measurements through VSM indicated an increase in the saturation magnetization ($M_S$) with the Zn/Ni and Cu/Ni content up to x=0.5 and a decrease for x>0.5. We have argued that this is due to an increase in the concentration of the $Fe^{+3}$ ions in the octahedral sites. This increase in $M_S$ is the highest for the case of $Ni_{0.5}Zn_{0.5}Fe_2O_4$ nanocrystals. All samples showed superparamagnetism behavior except $CuFe_2O_4$ nanocrystals.

We have also observed two broad emission bands in the PL spectra. These bands correspond to the presence of the indirect and direct band gaps for synthesized nanocrystals. Remarkably, the indirect band gap of Zn/Ni ferrite nanocrystals has shown an increase with the zinc content.

**Acknowledgements**

The authors acknowledge the Iranian Nanotechnology Initiative Council and the Alzahra University.

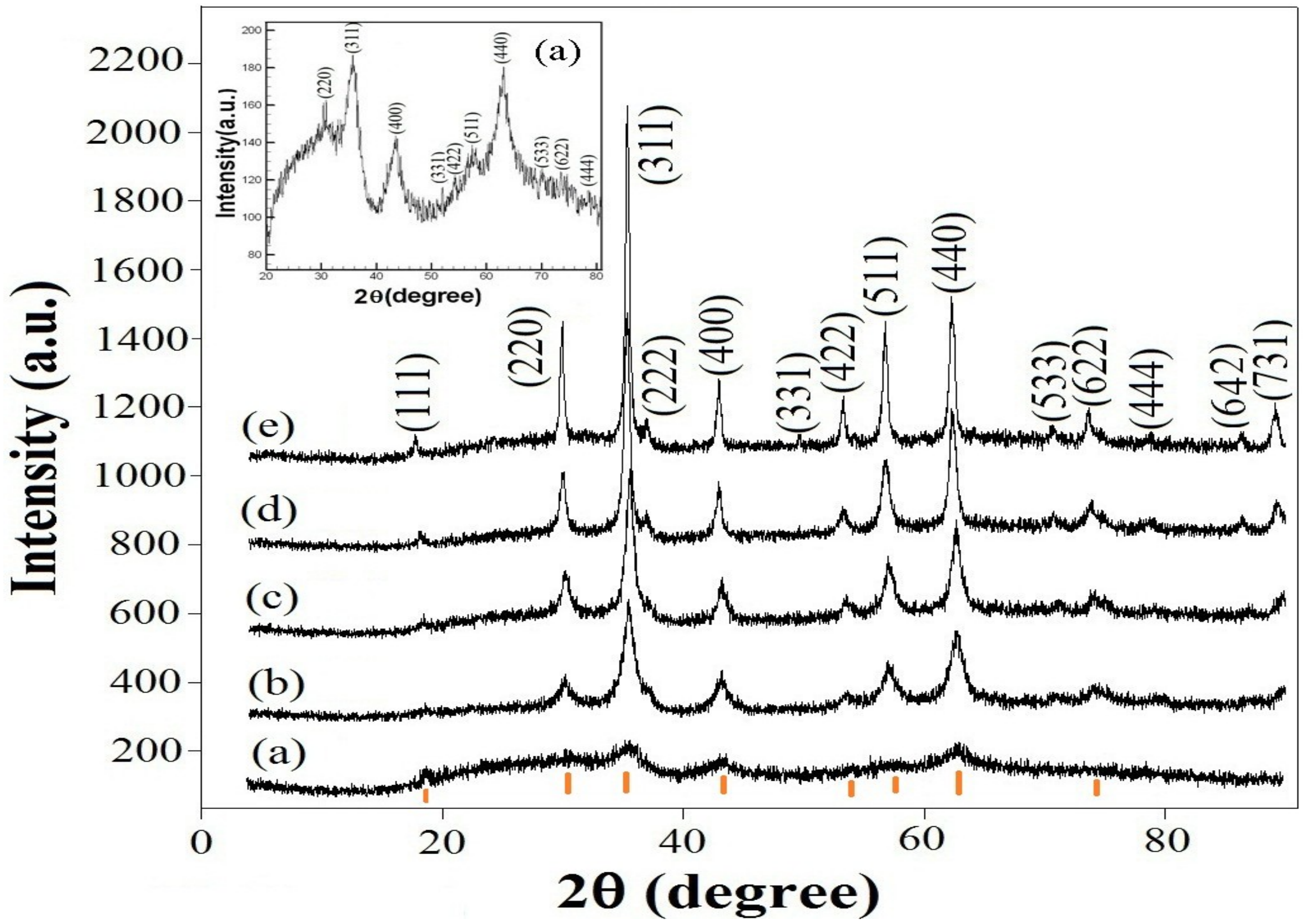

Fig. 1. XRD patterns of synthesized $Ni_{1-x}Zn_xFe_2O_4$ nanocrystals: (a) $NiFe_2O_4$, (b) $Ni_{0.7}Zn_{0.3}Fe_2O_4$ , (c) $Ni_{0.5}Zn_{0.5}Fe_2O_4$, (d) $Ni_{0.3}Zn_{0.7}Fe_2O_4$ , (e) $ZnFe_2O_4$ ( | : Bragg reflection positions)

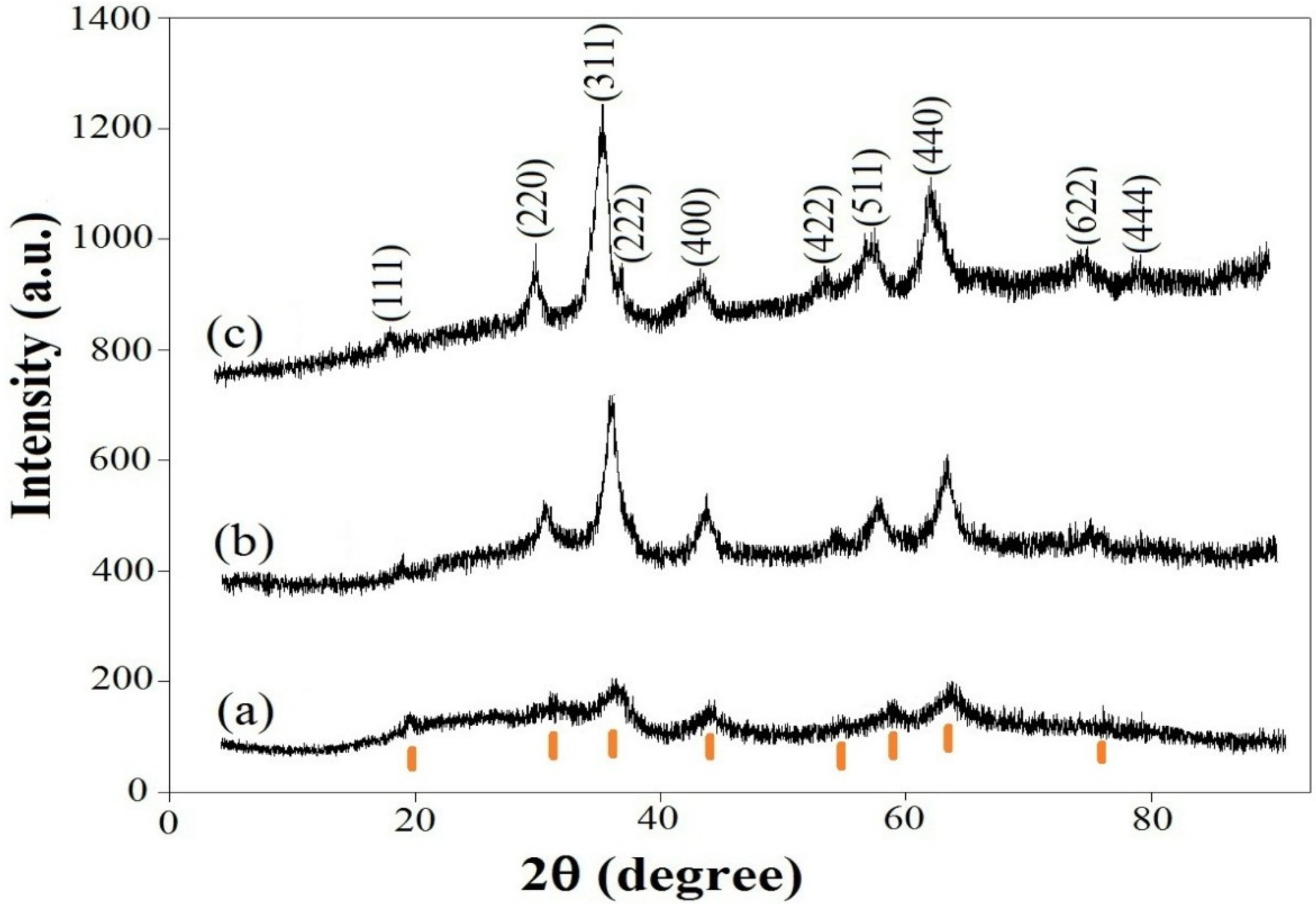

Fig. 2. XRD patterns of synthesized $Ni_{1-x}Cu_xFe_2O_4$ nanocrystals: (a) $NiFe_2O_4$, (b) $Ni_{0.5}Cu_{0.5}Fe_2O_4$, (c) $CuFe_2O_4$ ( | : Bragg reflection positions)

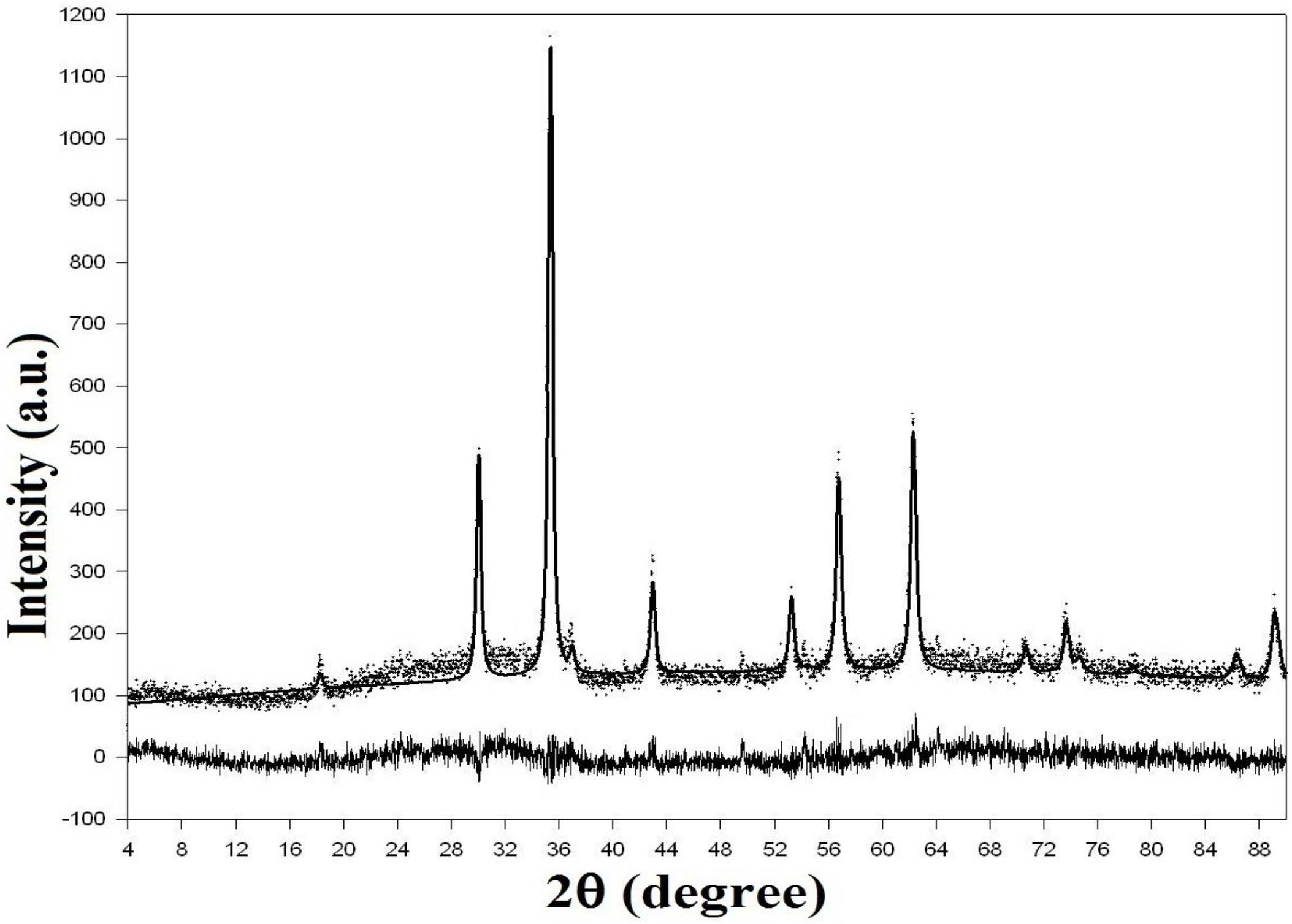

Fig. 3. XRD pattern refinements using MAUD software for $ZnFe_2O_4$ nanocrystals ( •: experimental data, upper solid line: calculated pattern, lower solid line: subtracted pattern)

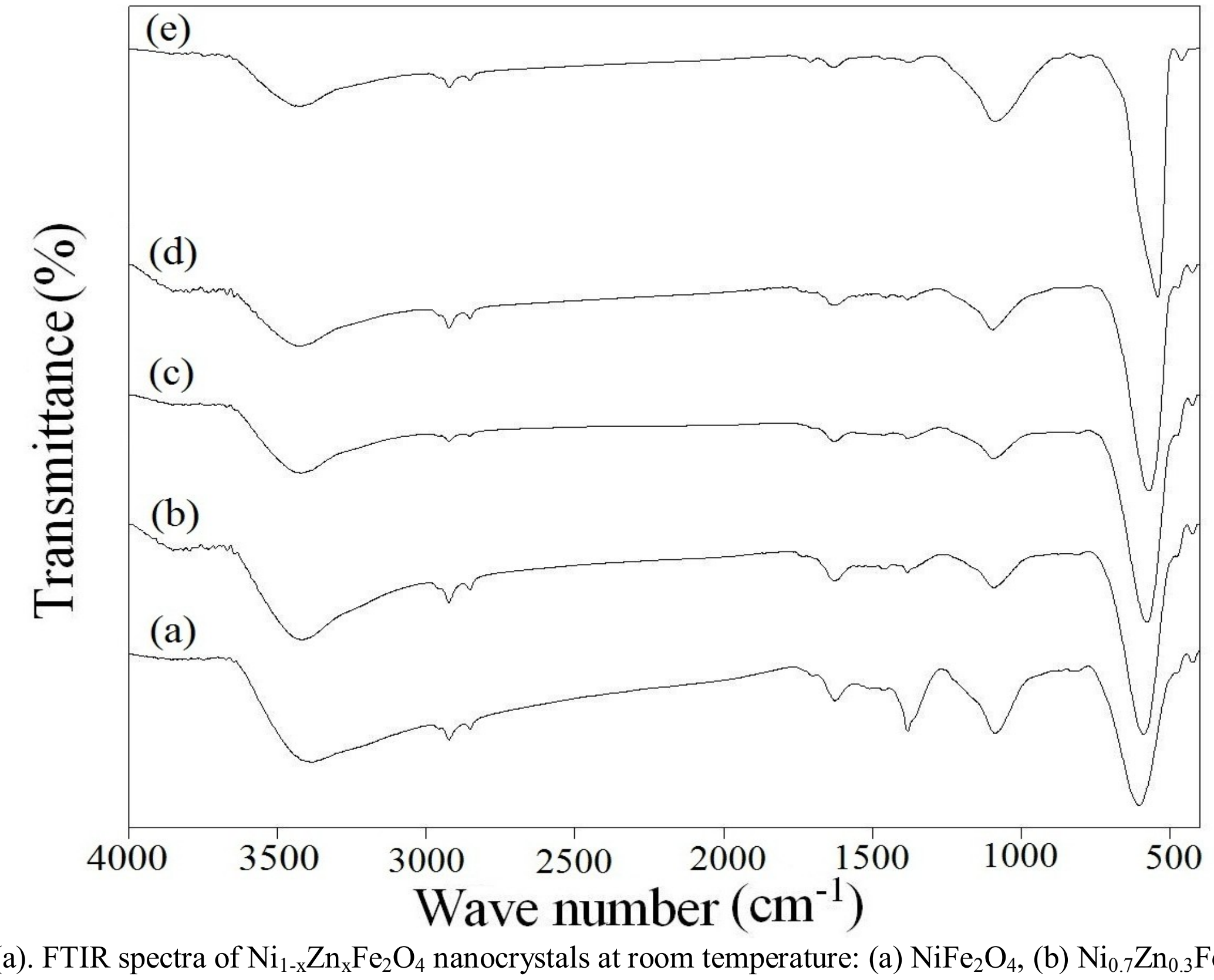

Fig. 4(a). FTIR spectra of $Ni_{1-x}Zn_xFe_2O_4$ nanocrystals at room temperature: (a) $NiFe_2O_4$, (b) $Ni_{0.7}Zn_{0.3}Fe_2O_4$ , (c) $Ni_{0.5}Zn_{0.5}Fe_2O_4$, (d) $Ni_{0.3}Zn_{0.7}Fe_2O_4$ , (e) $ZnFe_2O_4$

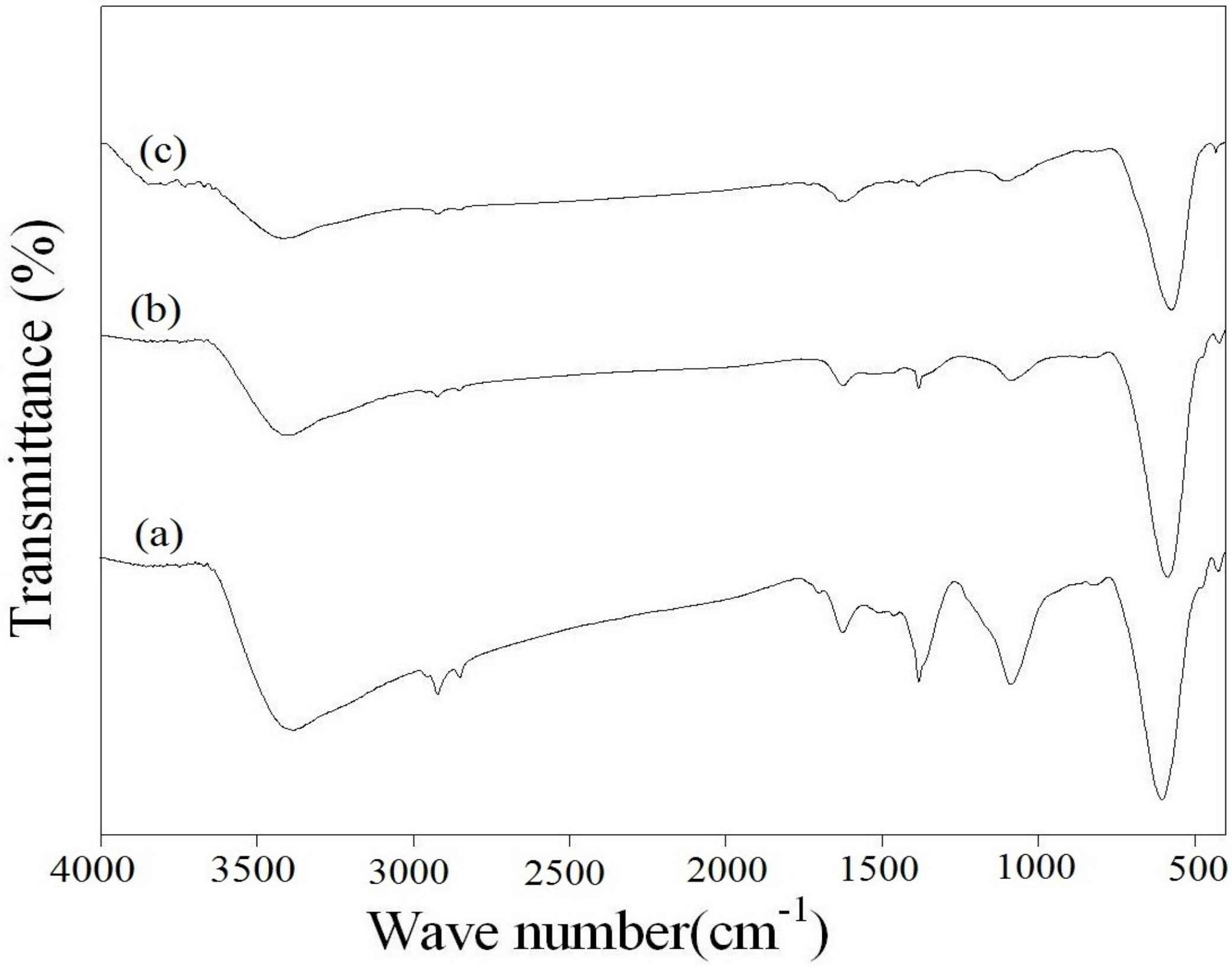

Fig. 4(b). FTIR spectra of $Ni_{1-x}Cu_xFe_2O_4$ nanocrystals at room temperature: (a) $NiFe_2O_4$, (b) $Ni_{0.5}Cu_{0.5}Fe_2O_4$, (c) $CuFe_2O_4$

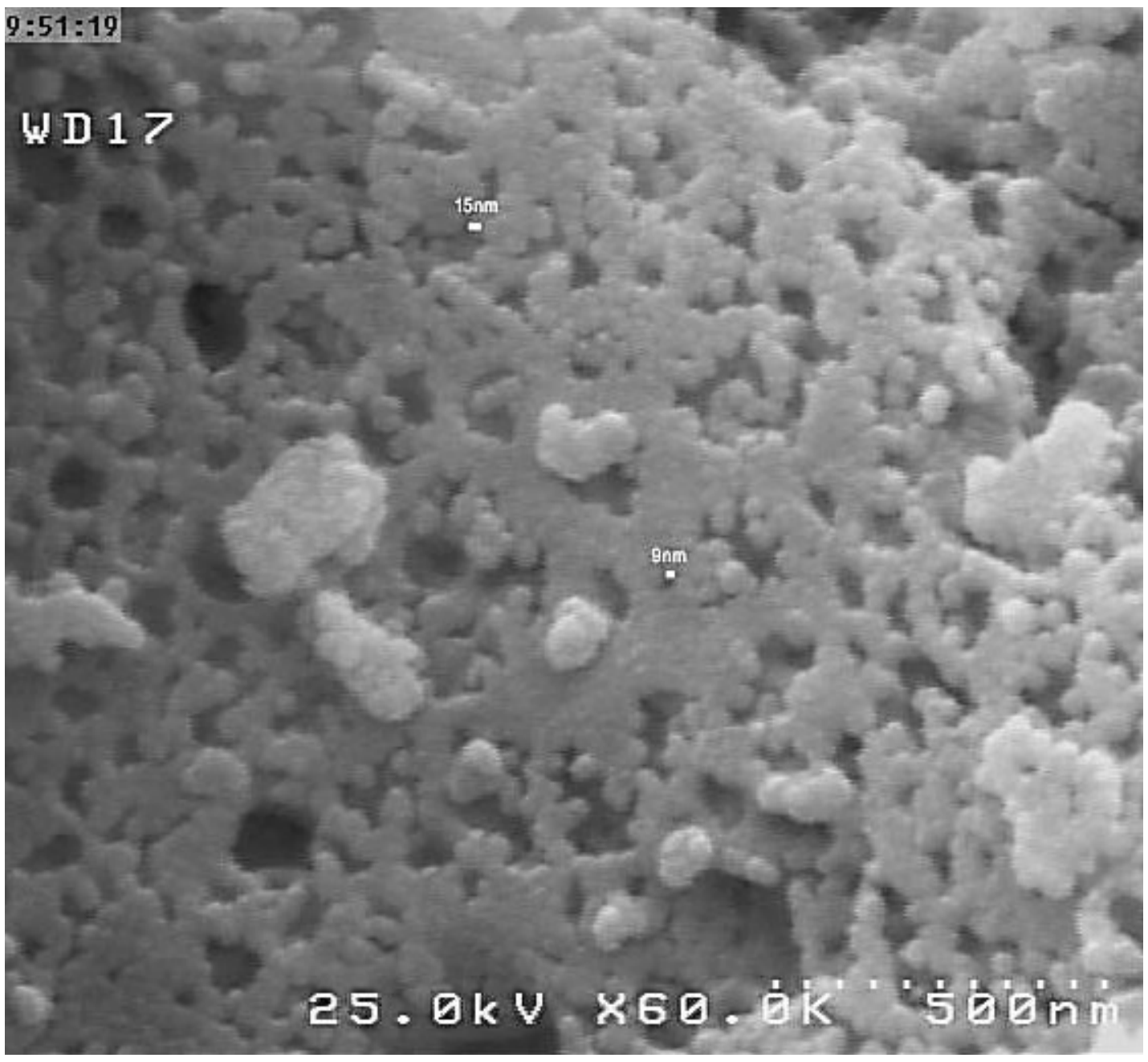

Fig. 5(a). FE-SEM image of synthesized $CuFe_2O_4$ nanocrystals

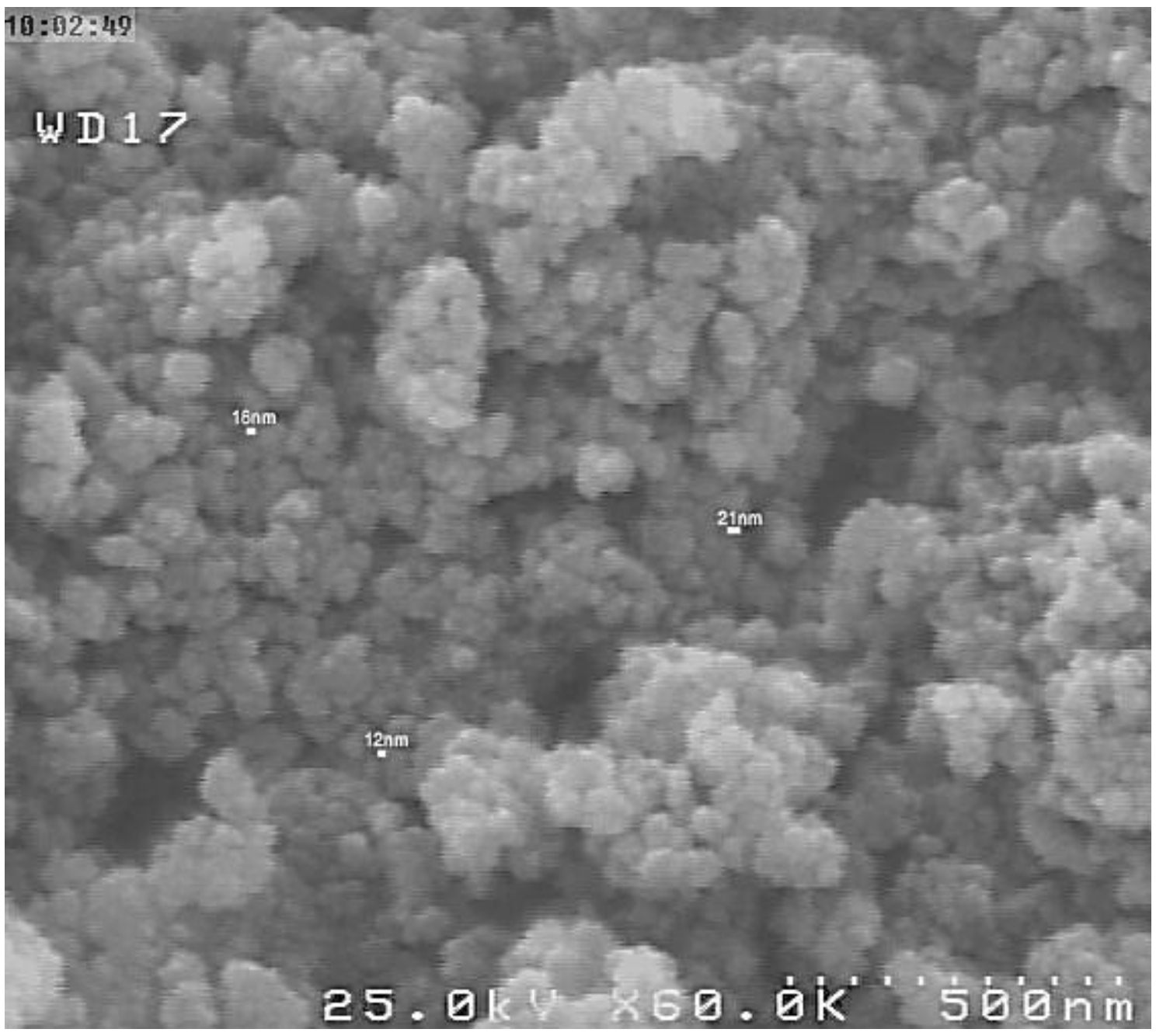

Fig. 5(b). FE-SEM image of synthesized $Ni_{0.5}Zn_{0.5}Fe_2O_4$ nanocrystals

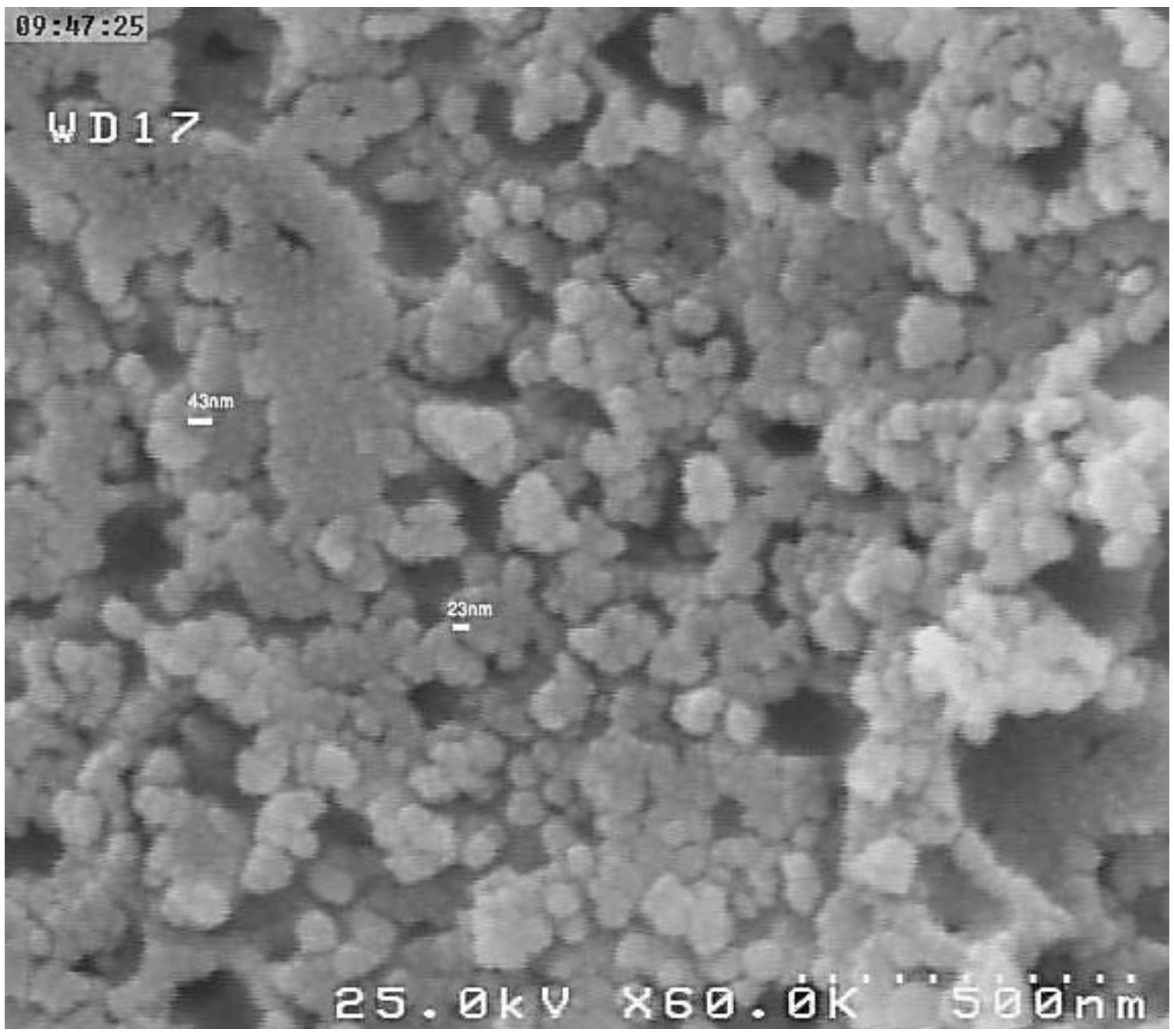

Fig. 5(c). FE-SEM image of synthesized $ZnFe_2O_4$ nanocrystals

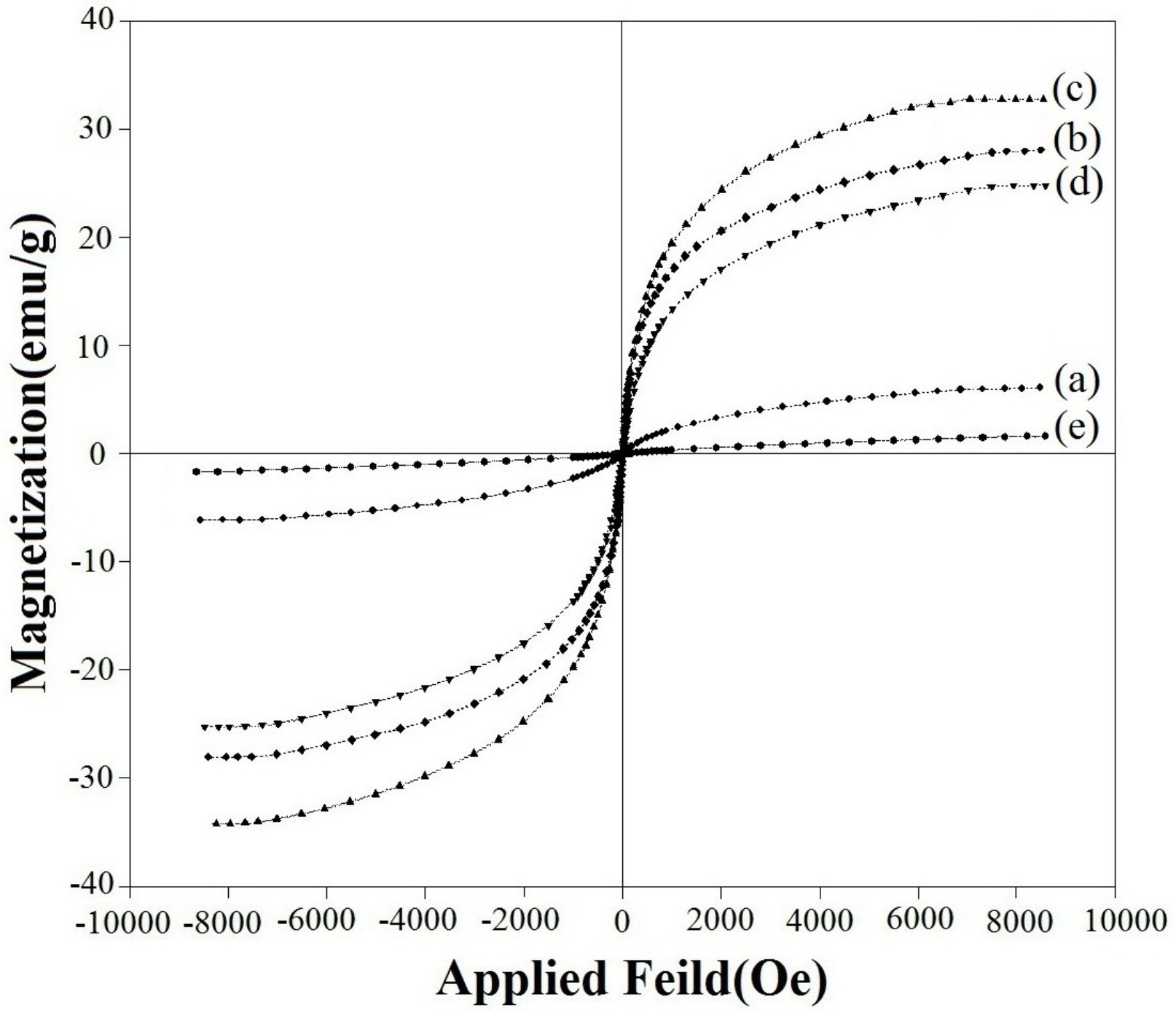

Fig. 6(a). Hysteresis curves of $Ni_{1-x}Zn_xFe_2O_4$ nanocrystals at room temperature: (a) $NiFe_2O_4$, (b) $Ni_{0.7}Zn_{0.3}Fe_2O_4$, (c) $Ni_{0.5}Zn_{0.5}Fe_2O_4$, (d) $Ni_{0.3}Zn_{0.7}Fe_2O_4$ , (e) $ZnFe_2O_4$

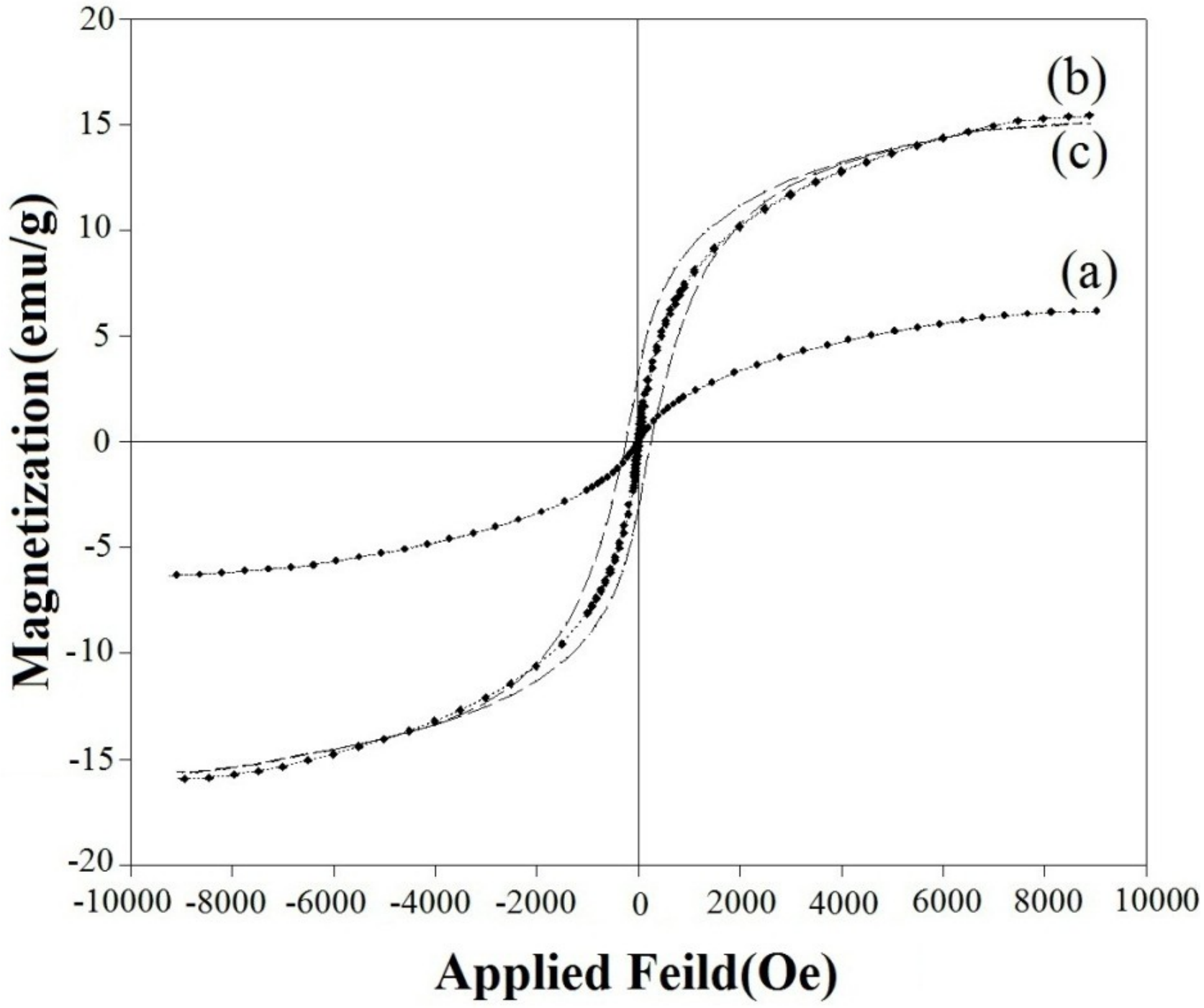

Fig. 6(b). Hysteresis curves of $Ni_{1-x}Cu_xFe_2O_4$ nanocrystals at room temperature: (a) $NiFe_2O_4$, (b) $Ni_{0.5}Cu_{0.5}Fe_2O_4$, (c) $CuFe_2O_4$

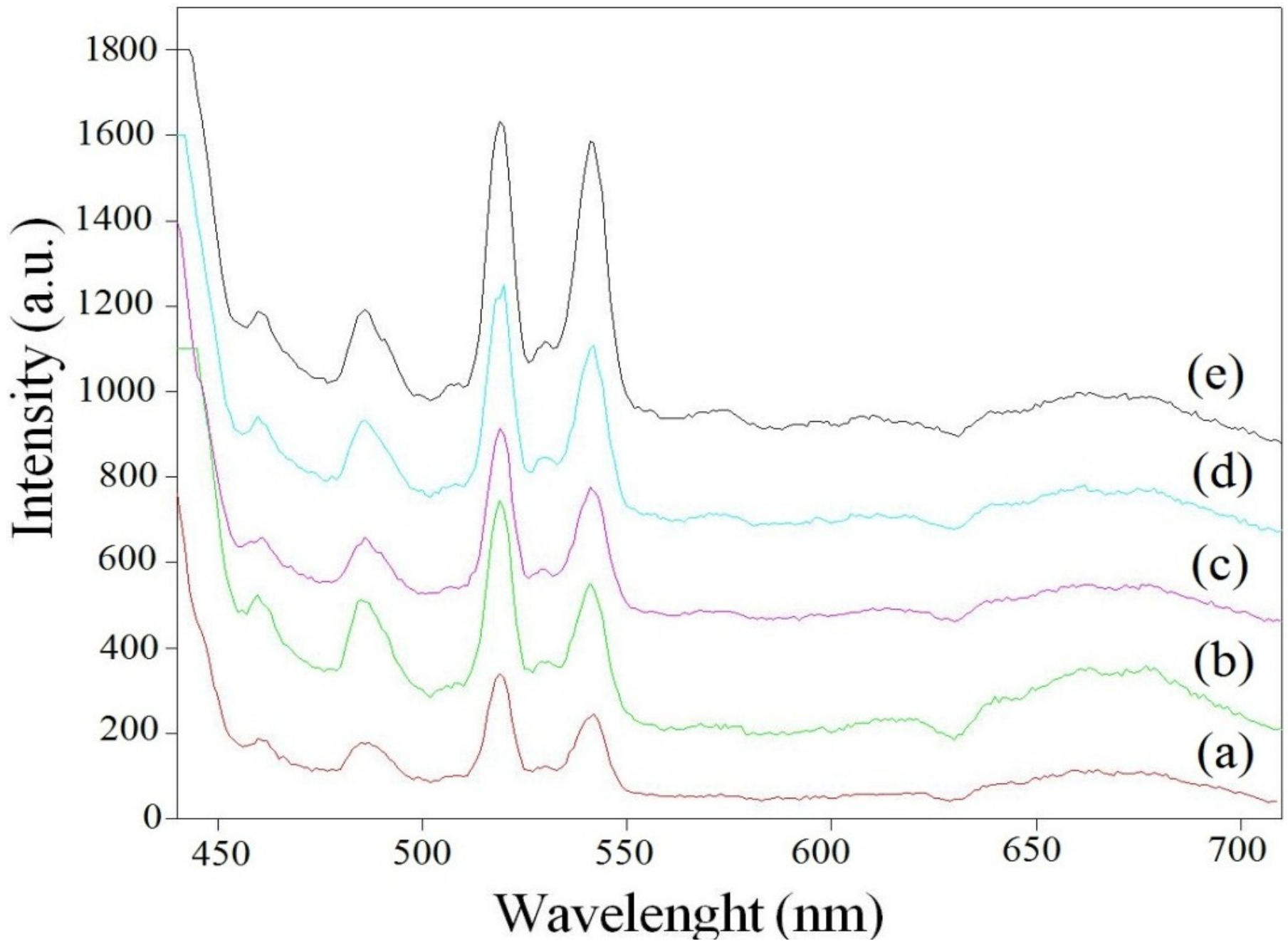

Fig.7 (a). Photoluminescence spectra of $Ni_{1-x}Zn_xFe_2O_4$ nanocrystals at room temperature: (a) $NiFe_2O_4$, (b) $Ni_{0.7}Zn_{0.3}Fe_2O_4$, (c) $Ni_{0.5}Zn_{0.5}Fe_2O_4$, (d) $Ni_{0.3}Zn_{0.7}Fe_2O_4$, (e) $ZnFe_2O_4$

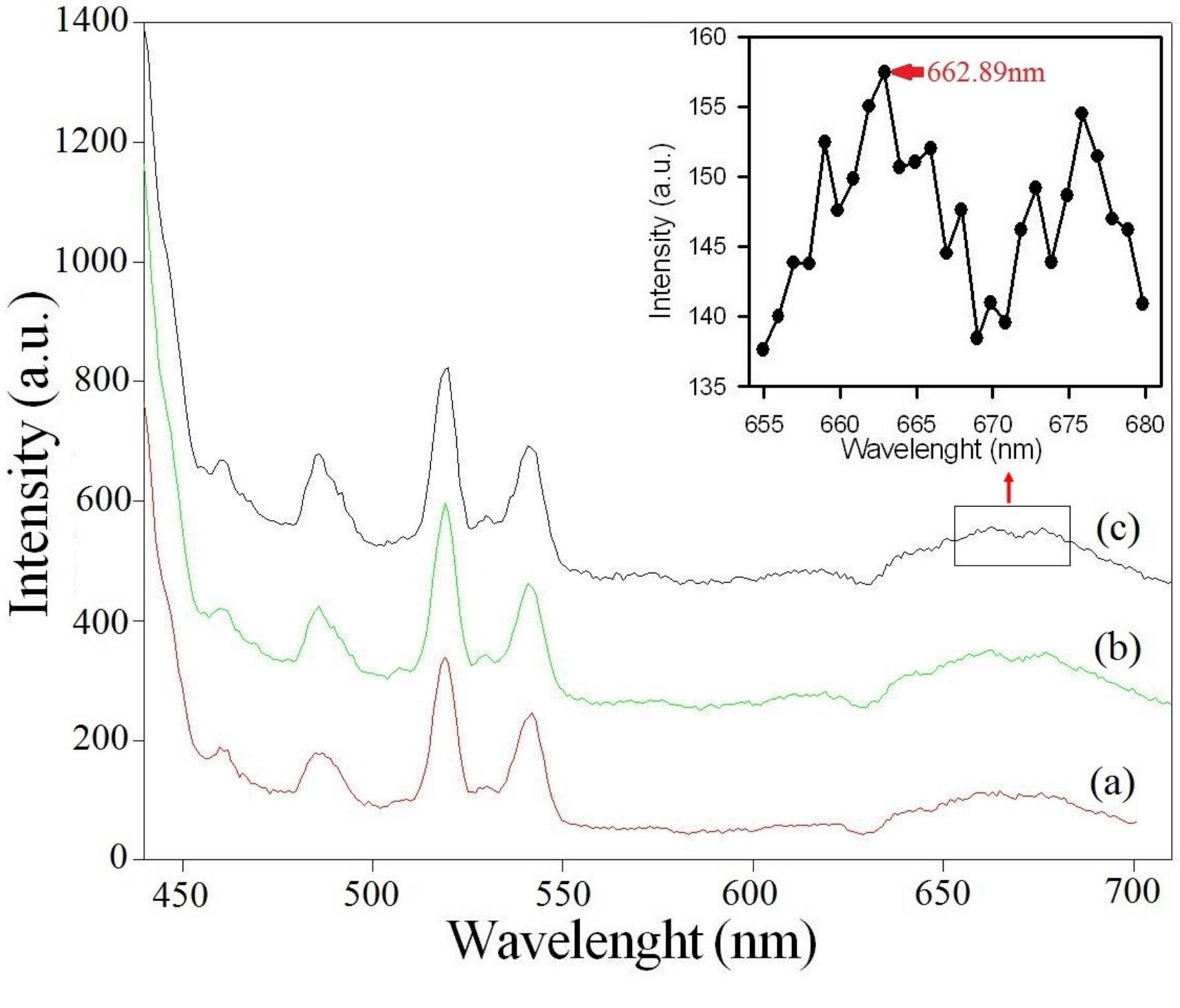

Fig. 7(b). Photoluminescence spectra of $Ni_{1-x}Cu_xFe_2O_4$ nanocrystals at room temperature: (a) $NiFe_2O_4$, (b) $Ni_{0.5}Cu_{0.5}Fe_2O_4$, (c) $CuFe_2O_4$

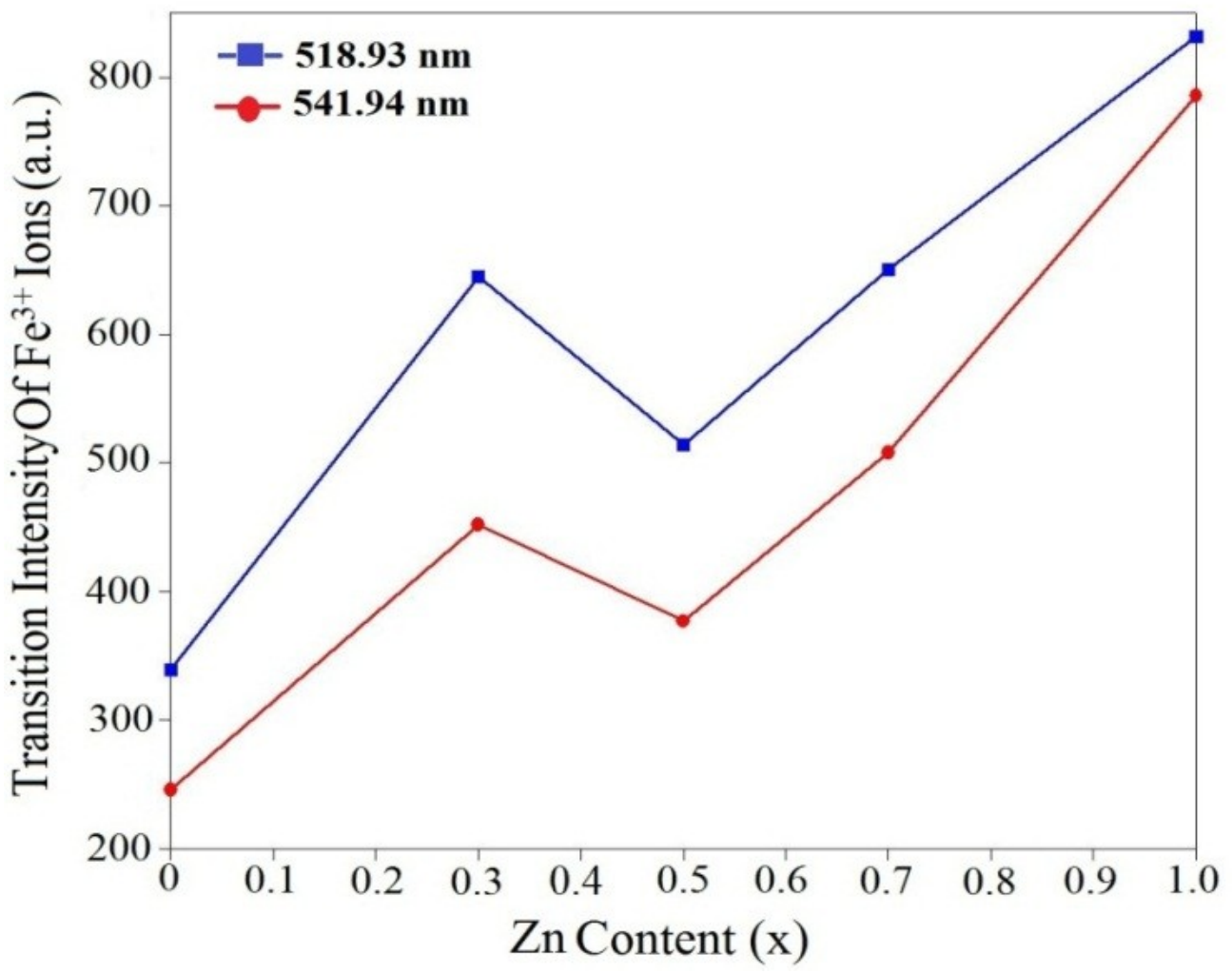

Fig. 8. Variation of the transition intensity of $Fe^{3+}$ ions versus the Zn content in $Ni_{1-x}Zn_xFe_2O_4$ nanocrystals

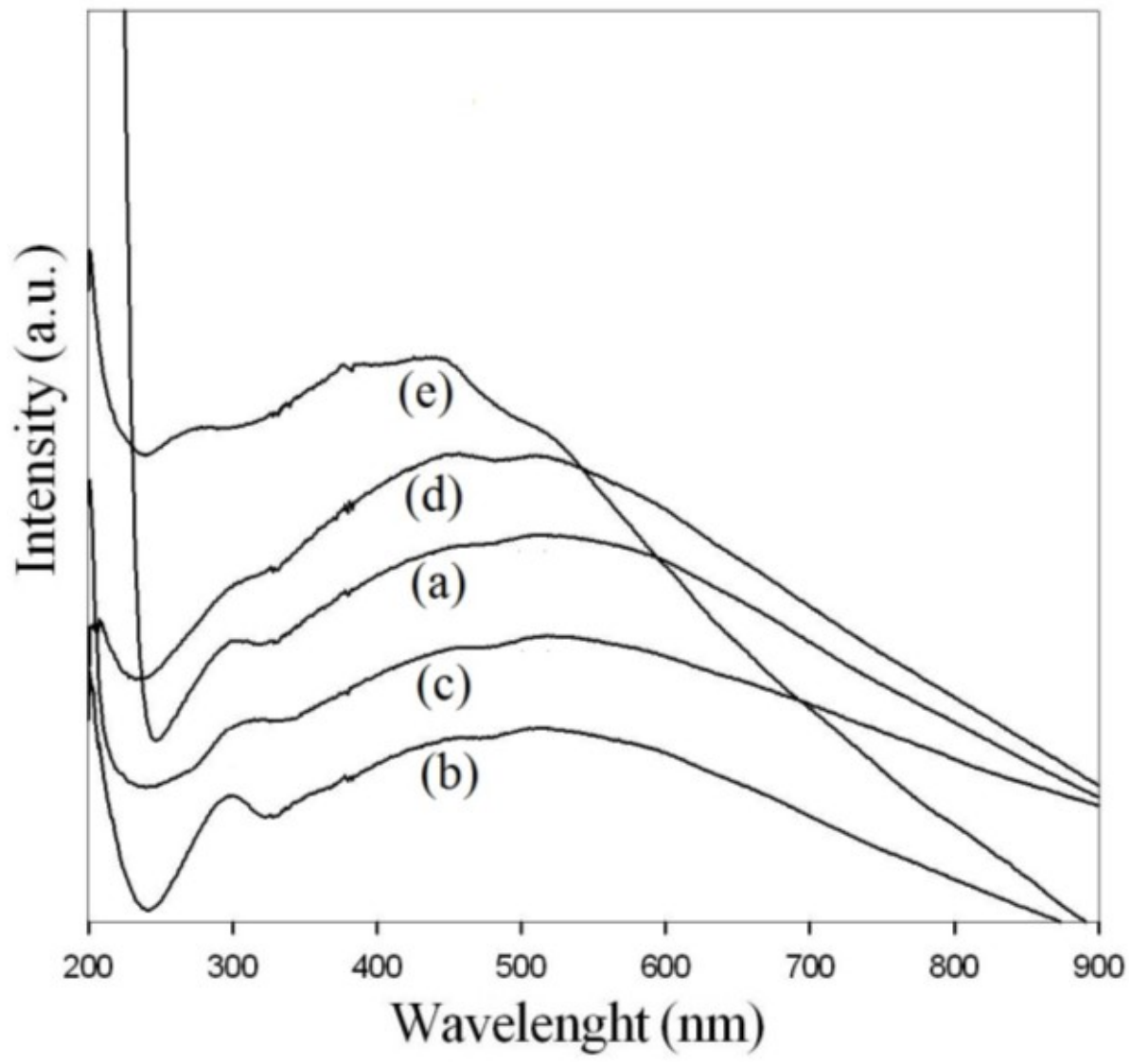

Fig. 9. UV-Vis spectra of $Ni_{1-x}Zn_xFe_2O_4$ nanocrystals at room temperature: (a) $NiFe_2O_4$, (b) $Ni_{0.7}Zn_{0.3}Fe_2O_4$, (c) $Ni_{0.5}Zn_{0.5}Fe_2O_4$, (d) $Ni_{0.3}Zn_{0.7}Fe_2O_4$, (e) $ZnFe_2O_4$

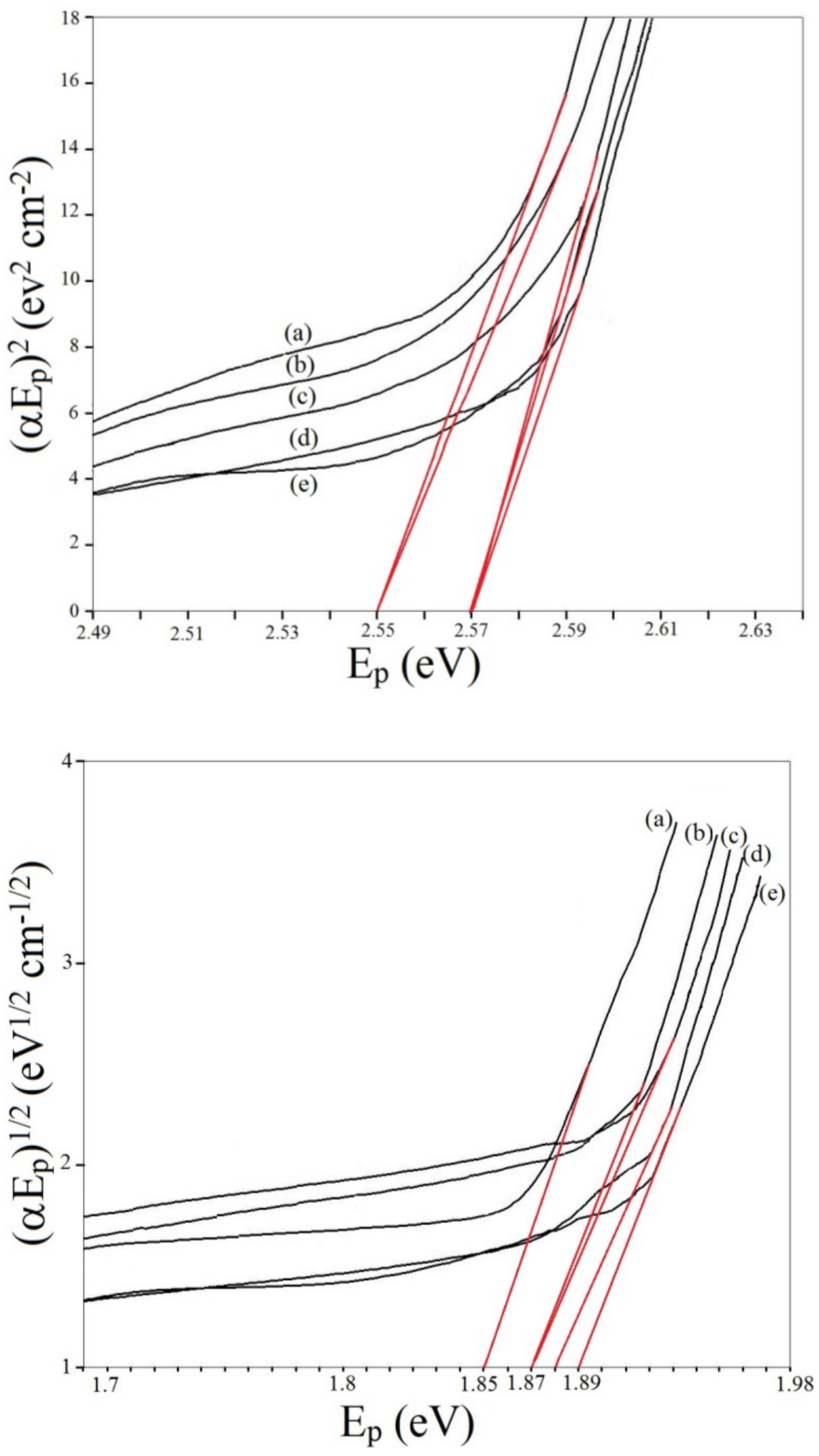

Fig 10. Plots of $(\alpha E_p)^2$ and $(\alpha E_p)^{1/2}$ versus $E_p$ for $Ni_{1-x}Zn_xFe_2O_4$ nanocrystals at room temperature: (a) $NiFe_2O_4$, (b) $Ni_{0.7}Zn_{0.3}Fe_2O_4$, (c) $Ni_{0.5}Zn_{0.5}Fe_2O_4$, (d) $Ni_{0.3}Zn_{0.7}Fe_2O_4$, (e) $ZnFe_2O_4$

| Zn content (x) | Composition | Cation distribution | Crystalline size (nm) | Lattice parameter (Å) | Saturation magnetization (emu/g) | Coercivity (Oe) |
|---|---|---|---|---|---|---|
| 0 | $NiFe_2O_4$ | $(Fe_{1.0}^{3+})_{tet}[Ni_{1.0}^{2+}, Fe_{1.0}^{3+}]_{oct}$ | 1.4 | 8.340 | 6.4 | 0.6 |
| 0.3 | $Ni_{0.7}Zn_{0.3}Fe_2O_4$ | $(Zn_{0.3}^{2+}, Fe_{0.7}^{3+})_{tet}[Ni_{0.7}^{2+}Fe_{1.3}^{3+}]_{oct}$ | 6.2±0.1 | 8.370 | 25.6 | 0.7 |
| 0.5 | $Ni_{0.5}Zn_{0.5}Fe_2O_4$ | $(Zn_{0.5}^{2+}, Fe_{0.5}^{3+})_{tet}[Ni_{0.5}^{2+}Fe_{1.5}^{3+}]_{oct}$ | 9.8 ± 0.1 | 8.389 | 34.8 | 0.9 |
| 0.7 | $Ni_{0.3}Zn_{0.7}Fe_2O_4$ | $(Zn_{0.7}^{2+}, Fe_{0.3}^{3+})_{tet}[Ni_{0.3}^{2+}Fe_{1.7}^{3+}]_{oct}$ | 13.5±0.2 | 8.400 | 28.4 | 1.0 |
| 1 | $ZnFe_2O_4$ | $(Zn_{1.0}^{2+})_{tet}[Fe_{1.0}^{3+}, Fe_{1.0}^{3+}]_{oct}$ | 19.6± 0.4 | 8.430 | 1.7 | 3.3 |

Table1. Crystalline size, lattice parameter, saturation magnetization, and coercivity of $Ni_{1-x}Zn_xFe_2O_4$ nanocrystals

| Cu content (x) | Composition | Cation distribution | Crystalline size (nm) | Lattice parameter (Å) | Saturation magnetization (emu/g) | Coercivity (Oe) |
|---|---|---|---|---|---|---|
| 0 | $NiFe_2O_4$ | $(Fe_{1.0}^{3+})_{tet}[Ni_{1.0}^{2+}, Fe_{1.0}^{3+}]_{oct}$ | 1.4 | 8.340 | 6.4 | 0.6 |
| 0.5 | $Ni_{0.5}Cu_{0.5}Fe_2O_4$ | $(Cu_{0.5}^{2+}, Fe_{0.5}^{3+})_{tet}[Ni_{0.5}^{2+}Fe_{1.5}^{3+}]_{oct}$ | 4.6± 0.1 | 8.355 | 16.2 | 2.3 |
| 1 | $CuFe_2O_4$ | $(Fe_{1.0}^{3+})_{tet}[Cu_{1.0}^{2+}, Fe_{1.0}^{3+}]_{oct}$ | 5.8± 0.1 | 8.370 | 14.9 | 168.2 |

Table2. Crystalline size, lattice parameter, saturation magnetization, and coercivity of $Ni_{1-x}Cu_xFe_2O_4$ nanocrystals

| Composition | Structural parameters before refinement | | | | | | Structural parameters after refinement | | | | | |
|---|---|---|---|---|---|---|---|---|---|---|---|---|
| | | Position | | | Occupancy | Quantity | | Position | | | Occupancy | **Quantity** |
| | Ions | X(Å) | Y(Å) | Z(Å) | | | Ions | X(Å) | Y(Å) | Z(Å) | | |
| **$NiFe_2O_4$** | $Fe^{3+}_{(tet)}$ | 0 | 0 | 0 | 1 | 8 | $Fe^{3+}_{(tet)}$ | $1.5\times10^{-7}\approx 0$ | $1.8\times10^{-7}\approx 0$ | $2.3\times10^{-7}\approx 0$ | 1.00002 | **8.00002** |
| | $Fe^{3+}_{(oct)}$ | 0.625 | 0.625 | 0.625 | 1 | 8 | $Fe^{3+}_{(oct)}$ | 0.62511 | 0.62540 | 0.62536 | 0.50003 | **8.00002** |
| | $Ni^{2+}_{(oct)}$ | 0.625 | 0.625 | 0.625 | 0.5 | 8 | $Ni^{2+}_{(oct)}$ | 0.62541 | 0.62544 | 0.62513 | 0.50002 | **8.00003** |
| | $O^{2-}$ | 0.3825 | 0.3825 | 0.3825 | 1 | 32 | $O^{2-}$ | 0.382502 | 0.382508 | 0.382504 | 1.00003 | **32.0002** |
| **$Ni_{0.7}Zn_{0.3}Fe_2O_4$** | $Fe^{3+}_{(tet)}$ | 0 | 0 | 0 | 0.7 | 5.6 | $Fe^{3+}_{(tet)}$ | $4.9\times10^{-6}\approx 0$ | $4.9\times10^{-6}\approx 0$ | $4.7\times10^{-6}\approx 0$ | 0.700018 | **5.60003** |
| | $Zn^{2+}_{(tet)}$ | 0 | 0 | 0 | 0.3 | 2.4 | $Zn^{2+}_{(tet)}$ | $6.2\times10^{-6}\approx 0$ | $5.5\times10^{-6}\approx 0$ | $3.5\times10^{-6}\approx 0$ | 0.300024 | **2.40003** |
| | $Fe^{3+}_{(oct)}$ | 0.625 | 0.625 | 0.625 | 0.65 | 10.4 | $Fe^{3+}_{(oct)}$ | 0.6248848 | 0.6249957 | 0.6250193 | 0.6500002 | **10.4001** |
| | $Ni^{2+}_{(oct)}$ | 0.625 | 0.625 | 0.625 | 0.35 | 5.6 | $Ni^{2+}_{(oct)}$ | 0.6248851 | 0.6250003 | 0.624773 | 0.3500001 | **5.60004** |
| | $O^{2-}$ | 0.3825 | 0.3825 | 0.3825 | 1 | 32 | $O^{2-}$ | 0.382521 | 0.382522 | 0.382515 | 1.000018 | **32.0006** |
| **$Ni_{0.5}Zn_{0.5}Fe_2O_4$** | $Fe^{3+}_{(tet)}$ | 0 | 0 | 0 | 0.5 | 4 | $Fe^{3+}_{(tet)}$ | $4.4\times10^{-6}\approx 0$ | $8.2\times10^{-6}\approx 0$ | $4.2\times10^{-6}\approx 0$ | 0.500008 | **4.00071** |
| | $Zn^{2+}_{(tet)}$ | 0 | 0 | 0 | 0.5 | 4 | $Zn^{2+}_{(tet)}$ | $9.4\times10^{-6}\approx 0$ | $3.4\times10^{-6}\approx 0$ | $5.9\times10^{-6}\approx 0$ | 0.500004 | **4.00002** |
| | $Fe^{3+}_{(oct)}$ | 0.625 | 0.625 | 0.625 | 0.75 | 12 | $Fe^{3+}_{(oct)}$ | 0.625183 | 0.625014 | 0.6250024 | 0.749999 | **11.9999** |
| | $Ni^{2+}_{(oct)}$ | 0.625 | 0.625 | 0.625 | 0.25 | 4 | $Ni^{2+}_{(oct)}$ | 0.624531 | 0.625434 | 0.6250033 | 0.249999 | **4.00002** |
| | $O^{2-}$ | 0.3825 | 0.3825 | 0.3825 | 1 | 32 | $O^{2-}$ | 0.3824999 | 0.3825001 | 0.3825001 | 1.00001 | **32.0003** |
| **$Ni_{0.3}Zn_{0.7}Fe_2O_4$** | $Fe^{3+}_{(tet)}$ | 0 | 0 | 0 | 0.3 | 2.4 | $Fe^{3+}_{(tet)}$ | $3.4\times10^{-6}\approx 0$ | $7.6\times10^{-6}\approx 0$ | $6.3\times10^{-6}\approx 0$ | 0.300002 | **2.40003** |
| | $Zn^{2+}_{(tet)}$ | 0 | 0 | 0 | 0.7 | 5.6 | $Zn^{2+}_{(tet)}$ | $2.2\times10^{-6}\approx 0$ | $7.7\times10^{-6}\approx 0$ | $6.3\times10^{-6}\approx 0$ | 0.699993 | **5.58996** |
| | $Fe^{3+}_{(oct)}$ | 0.625 | 0.625 | 0.625 | 0.85 | 13.6 | $Fe^{3+}_{(oct)}$ | 0.624888 | 0.625001 | 0.625051 | 0.850004 | **13.6004** |
| | $Ni^{2+}_{(oct)}$ | 0.625 | 0.625 | 0.625 | 0.15 | 2.4 | $Ni^{2+}_{(oct)}$ | 0.625004 | 0.62513 | 0.624859 | 0.150003 | **2.40057** |
| | $O^{2-}$ | 0.3825 | 0.3825 | 0.3825 | 1 | 32 | $O^{2-}$ | 0.382488 | 0.382511 | 0.3825055 | 1.000002 | **32.0006** |
| **$ZnFe_2O_4$** | $Zn^{2+}_{(tet)}$ | 0 | 0 | 0 | 1 | 8 | $Zn^{2+}_{(tet)}$ | $1.1\times10^{-6}\approx 0$ | $1.1\times10^{-6}\approx 0$ | $1.1\times10^{-6}\approx 0$ | 1.00008 | **8.00065** |
| | $Fe^{3+}_{(oct)}$ | 0.625 | 0.625 | 0.625 | 1 | 16 | $Fe^{3+}_{(oct)}$ | 0.625001 | 0.65002 | 0.625302 | 1.00001 | **16.0001** |
| | $O^{2-}$ | 0.3825 | 0.3825 | 0.3825 | 1 | 32 | $O^{2-}$ | 0.382504 | 0.382515 | 0.382505 | 0.999999 | **31.9999** |
| **$Ni_{0.5}Cu_{0.5}Fe_2O_4$** | $Fe^{3+}_{(tet)}$ | 0 | 0 | 0 | 0.5 | 4 | $Fe^{3+}_{(tet)}$ | $1.3\times10^{-6}\approx 0$ | $3.1\times10^{-6}\approx 0$ | $3.7\times10^{-6}\approx 0$ | 0.50001 | **4.00001** |
| | $Cu^{2+}_{(tet)}$ | 0 | 0 | 0 | 0.5 | 4 | $Cu^{2+}_{(tet)}$ | $1.1\times10^{-6}\approx 0$ | $2.5\times10^{-6}\approx 0$ | $3.1\times10^{-6}\approx 0$ | 0.50002 | **4.00002** |
| | $Fe^{3+}_{(oct)}$ | 0.625 | 0.625 | 0.625 | 0.75 | 12 | $Fe^{3+}_{(oct)}$ | 0.625183 | 0.625014 | 0.6250024 | 0.74999 | **11.9999** |
| | $Ni^{2+}_{(oct)}$ | 0.625 | 0.625 | 0.625 | 0.25 | 4 | $Ni^{2+}_{(oct)}$ | 0.625031 | 0.625434 | 0.6250033 | 0.24999 | **3.99999** |
| | $O^{2-}$ | 0.3825 | 0.3825 | 0.3825 | 1 | 32 | $O^{2-}$ | 0.382499 | 0.3825001 | 0.3825012 | 1.000003 | **32.0001** |
| **$CuFe_2O_4$** | $Fe^{3+}_{(tet)}$ | 0 | 0 | 0 | 1 | 8 | $Fe^{3+}_{(tet)}$ | $1.6\times10^{-7}\approx 0$ | $4.2\times10^{-7}\approx 0$ | $1.8\times10^{-7}\approx 0$ | 1.00002 | **8.0002** |
| | $Fe^{3+}_{(oct)}$ | 0.625 | 0.625 | 0.625 | 0.5 | 8 | $Fe^{3+}_{(oct)}$ | 0.62541 | 0.62548 | 0.62566 | 0.50001 | **8.00001** |
| | $Cu^{2+}_{(oct)}$ | 0.625 | 0.625 | 0.625 | 0.5 | 8 | $Cu^{2+}_{(oct)}$ | 0.62513 | 0.62573 | 0.62532 | 0.49999 | **7.9999** |
| | $O^{2-}$ | 0.3825 | 0.3825 | 0.3825 | 1 | 32 | $O^{2-}$ | 0.382504 | 0.382515 | 0.382505 | 0.999999 | **31.9999** |

Table 3. Structural parameters before and after refinement from MAUD analysis for $Ni_{1-x}M_xFe_2O_4$ nanocrystals

| composition | $R_{wp}$ | $R_p$ | $R_{exp}$ | S |
|---|---|---|---|---|
| $NiFe_2O_4$ | 0.118 | 0.110 | 0.095 | 1.24 |
| $Ni_{0.7}Zn_{0.3}Fe_2O_4$ | 0.131 | 0.103 | 0.103 | 1.27 |
| $Ni_{0.5}Zn_{0.5}Fe_2O_4$ | 0.113 | 0.108 | 0.098 | 1.15 |
| $Ni_{0.3}Zn_{0.7}Fe_2O_4$ | 0.114 | 0.108 | 0.101 | 1.13 |
| $ZnFe_2O_4$ | 0.117 | 0.102 | 0.097 | 1.20 |
| $Ni_{0.5}Cu_{0.5}Fe_2O_4$ | 0.104 | 0.083 | 0.084 | 1.22 |
| $CuFe_2O_4$ | 0.089 | 0.069 | 0.073 | 1.21 |

Table 4. The parameters for the calculation of pattern fitness for $Ni_{1-x}M_xFe_2O_4$ nanocrystals

| Zn content (x) | Composition | FTIR frequency bands ($cm^{-1}$) | | Indirect band gap (eV) | Direct band gap (eV) |
|---|---|---|---|---|---|
| | | A- site (tet) $\upsilon_1$ | B-site (oct) $\upsilon_2$ | | |
| **0** | $NiFe_2O_4$ | 604.33 | 425.66 | 1.85 | 2.55 |
| **0.3** | $Ni_{0.7}Zn_{0.3}Fe_2O_4$ | 590.21 | 426.11 | 1.87 | 2.56 |
| **0.5** | $Ni_{0.5}Zn_{0.5}Fe_2O_4$ | 578.28 | 426.60 | 1.87 | 2.56 |
| **0.7** | $Ni_{0.3}Zn_{0.7}Fe_2O_4$ | 571.55 | 426.63 | 1.88 | 2.56 |
| **1** | $ZnFe_2O_4$ | 542.37 | 463.14 | 1.90 | 2.56 |

Table5. FTIR absorption band frequencies and optical parameter of $Ni_{1-x}Zn_xFe_2O_4$ nanocrystals

| Cu content (x) | Composition | FTIR frequency bands ($cm^{-1}$) | | | Indirect band gap (eV) | Direct band gap (eV) |
|---|---|---|---|---|---|---|
| | | A- site (tet) | B-site (oct) | | | |
| | | $\upsilon_1$ | $\upsilon_2$ | $\upsilon'_2$ | | |
| **0** | $NiFe_2O_4$ | 604.33 | 425.66 | – | 1.85 | 2.55 |
| **0.5** | $Ni_{0.5}Cu_{0.5}Fe_2O_4$ | 574.69 | 428.13 | – | 1.87 | 2.55 |
| **1** | $CuFe_2O_4$ | 585.86 | 422.22 | 416.61 | 1.87 | 2.55 |

Table6. FTIR absorption band frequencies and optical parameter of $Ni_{1-x}Cu_xFe_2O_4$ nanocrystals

**Shahbaz Tehrani, Fatemeh**

**B. Sc.**

Alzahra University, Tehran 19938, Iran

**M. Sc. student**

Alzahra University, Tehran 19938, Iran

**E-mail:** tehrani66@gmail.com

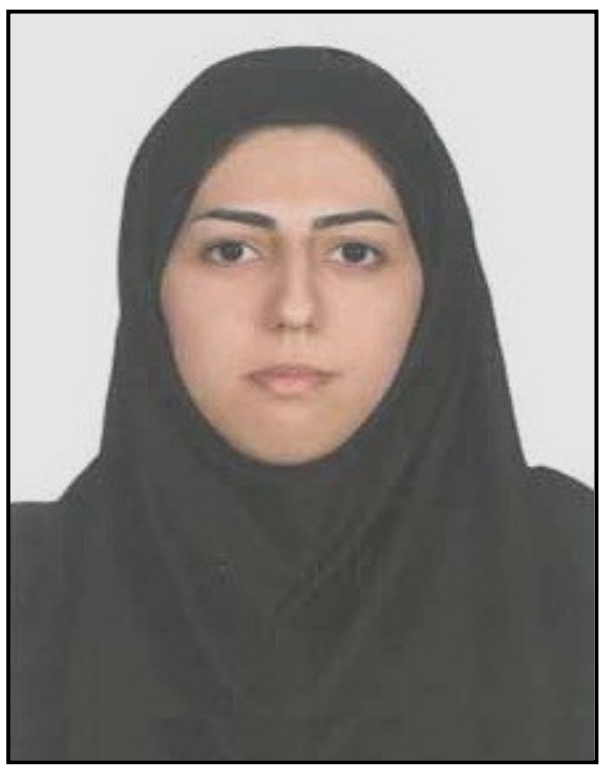

**Research interests:**

Ferrits, Nanocatalysts, Magnetic Nanoparticles, Carbon Nanotubes

**Corresponding author: Daadmehr, Vahid**

Director, Magnet & Superconducting Research Laboratory

Associate Professor of Physics

Department of Physics, Alzahra University, Tehran 19938, Iran

**Tel:** (+98 21) 85692640 / (+98) 912608 9714

**Fax:** (+98 21) 88047861

**E-mail:** daadmehr@alzahra.ac.ir

**Web:** http:// www.alzahra.ac.ir/daadmehr/

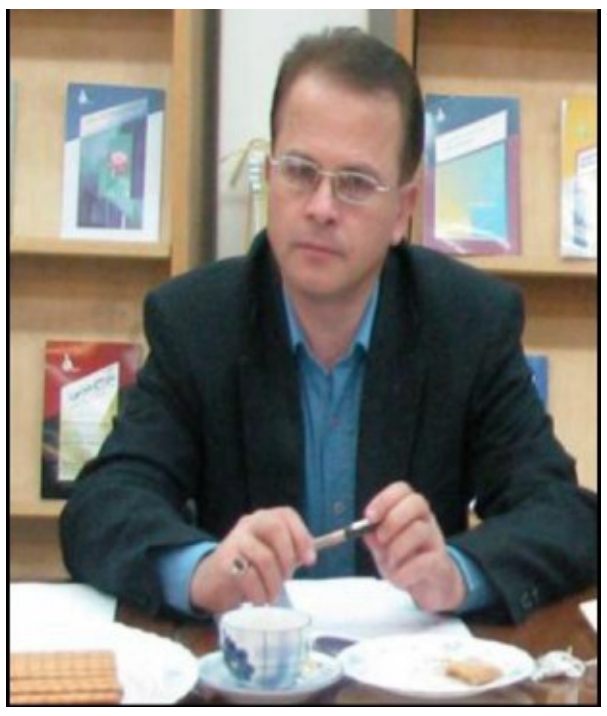

**Research interests:**

Condensed matter physics, Material science, nanoferrites, Carbon Nanostructures, High Temperature Superconductors: Electrical properties, Nanocrystals, Y-based cuprates.

**Rezakhani, Ali T.**

Assistant Professor of Physics

Department of Physics, Sharif University of Technology, Tehran, Iran

**Tel:** (+98 21) 6616 4523

**Fax:** (+98 21) 6602 2711

**E-mail:** rezakhani@sharif.edu

**Personal Page:** http://sharif.edu/~rezakhani

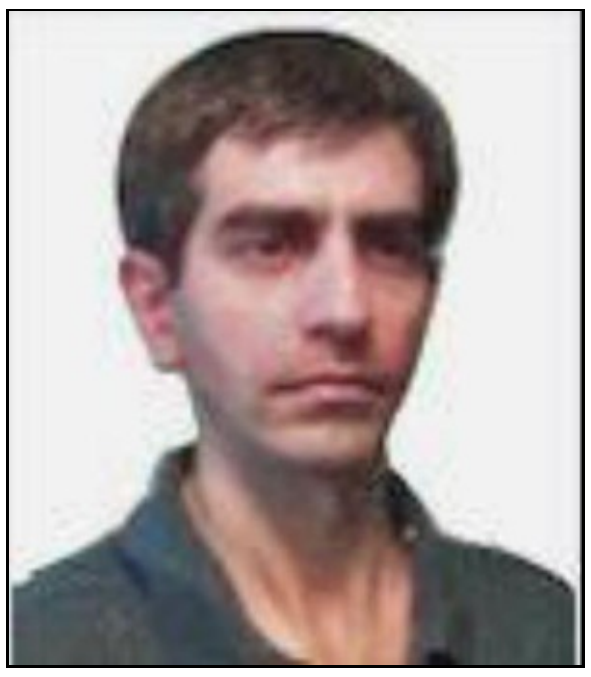

**Research interests:**

Quantum Information Science, Quantum Computation, Dynamics of Open Quantum Systems, Condensed matter-Superconductivity.

**Hosseini Akbarnejad, Razieh**

**B. Sc.**

Tehran University, Tehran 19938, Iran

**M. Sc. student**

Alzahra University, Tehran 19938, Iran

**E-mail:** rh.akbarnejad@gmail.com

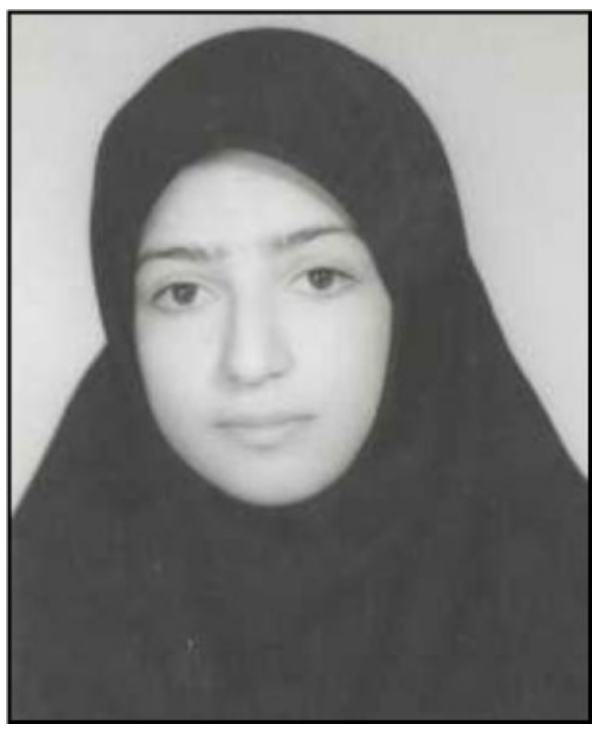

**Research interests:**

Carbon Nanotubes, Nanocatalysts, Magnetic Nanoparticles, Ferrites

**Gholipour, Somayyeh**

**B. Sc.**

Alzahra University, Tehran 19938, Iran

**M. Sc. student**

Alzahra University, Tehran 19938, Iran

**Email:** gholipour65@gmail.com

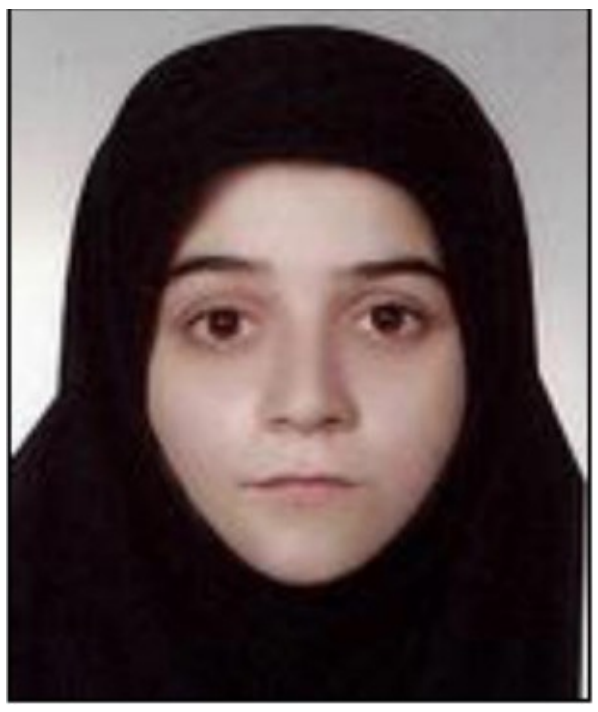

**Research interest:**

Ferrits, Nanocatalysts, High Temperature Superconductors: Electrical Properties, Nanocrystals, Y-based cuprates.